%% file: RJwrapper.tex
\documentclass{article}

\usepackage{arxiv}

\usepackage[utf8]{inputenc} 
\usepackage[T1]{fontenc}    
\usepackage{hyperref}       
\usepackage{url}            
\usepackage{booktabs}       
\usepackage{amsfonts}       
\usepackage{nicefrac}       
\usepackage{microtype}      
\usepackage{lipsum}		
\usepackage{graphicx}
\usepackage{natbib}
\usepackage{doi}
\usepackage{amsmath,amssymb,array}
\usepackage{amsthm}
\usepackage{cleveref}
\newtheorem{proposition}{Proposition}
\newtheorem{remark}{Remark}
\newtheorem*{Example*}{Example}

\usepackage{enumitem}

\usepackage{booktabs}

\graphicspath{{figures/}} 
\title{{\tt Splinets} -- Orthogonal Splines and FDA for the Classification Problem}

\date{} 					

\author{Rani Basna \thanks{All authors contributed equally} \\
        Department of Statistics, Lund University, Sweden,\\
	\texttt{rani.basna@stat.lu.se} \\
	\And
	Hiba Nassar\\
	Cognitive Systems, Department of Applied Mathematics and Computer Science,\\
        Technical University of Denmark, Denmark\\
	\texttt{hibna@dtu.dk} \\
	 \AND
	Krzysztof Podg\'orski \\
	Department of Statistics, Lund University, Sweden \\
	 \texttt{ Krzysztof.Podgorski@stat.lu.se} \\
}

\begin{document}
\maketitle
\begin{abstract}
This study introduces an efficient workflow for functional data analysis in classification problems, utilizing advanced orthogonal spline bases. The methodology is based on the flexible \href{https://cran.r-project.org/web/packages/Splinets/index.html}{\tt Splinets} package, featuring a novel spline representation designed for enhanced data efficiency. Several innovative features contribute to this efficiency: 1)~{\em Utilization of Orthonormal Spline Bases} -- The workflow incorporates recently introduced orthonormal spline bases, known as splinets, ensuring a robust foundation for data analysis; 2)~{\em Consideration of Spline Support Sets} -- The proposed spline object representation accounts for spline support sets, which refines the accuracy of data representation; 3)~{\em Data-Driven Knot Selection} -- The workflow employs data-driven knot selection techniques for spline construction, optimizing the overall analysis process.

To illustrate the effectiveness of this approach, we applied the workflow to the Fashion MINST dataset, a collection of two-dimensional images. We demonstrate the classification process and highlight significant efficiency gains. Particularly noteworthy are the improvements that can be achieved through the 2D generalization of our methodology, especially in scenarios where data sparsity and dimension reduction are critical factors.
A key advantage of our workflow is the projection operation into the space of splines with arbitrarily chosen knots, allowing for versatile functional data analysis associated with classification problems.

Moreover, this study explores some features of the \href{https://cran.r-project.org/web/packages/Splinets/index.html}{\tt Splinets} package suited for functional data analysis. The algebra and calculus of splines use Taylor expansions at the knots within the support sets. Various orthonormalization techniques for $B$-splines are implemented, including the highly recommended dyadic method, which leads to the creation of {\it splinets}. Importantly, the locality of $B$-splines concerning support sets is preserved in the corresponding {\it splinet}. Using this locality, along with implemented algorithms, provides a powerful computational tool for functional data analysis.
\end{abstract}


\input{RJtemplate}

\end{document}

%% file: RJtemplate.tex
\section{Introduction}
\label{sec:intro}
In functional data analysis (FDA), it is often desired that functions considered are continuous or even differentiable up to a certain order. 
From this perspective, spline functions given over a set of knots form convenient finite-dimensional functional spaces.  
Many \textbf{\textsf{R}}-packages that handle splines, see \cite{Perperoglou:2019aa}. 
However, none of them consistently treat splines as elements of functional spaces with explicitly evaluated orthogonal bases. 
The recent package, {\href{https://cran.r-project.org/web/packages/Splinets/index.html}{\tt Splinets}},  approaches splines exactly like this by using an object that represents a set of functional splines and provides efficient orthogonal bases as such objects.
This focus on the functional form of splines and functional analysis approach to them makes {\href{https://cran.r-project.org/web/packages/Splinets/index.html}{\tt Splinets}} particularly suitable for functional data analysis.  
Care was taken to make this functional treatment efficient by first accounting for support sets of spline, then using efficient and novel orthogonal bases. 

While the locality of the representation, expressed by the support sets, makes the package suitable for the analysis of sparse data, its most important feature is the orthogonal spline bases. Standard $B$-bases, \cite{Boor1978APG, schumaker2007spline}, are evaluated, and efficiency is achieved by explicitly using the support in implemented algebra and calculus of splines.  However, there is a problem with the $B$-splines bases that introduces a computational burden when decomposing a function for functional data analysis: the $B$-splines are not orthogonal.
Since any basis in a Hilbert space can be orthogonalized, it is to the point of considering orthonormalization of the $B$-splines, \cite{nguyen2015construction,goodman2003class,cho2005class}. In \cite{LIU2022}, a new natural orthogonalization method for the B-splines has been introduced and an efficient dyadic algorithm has been proposed for this purpose.  The orthogonalization was argued to be the most appropriate since it, firstly, preserves most from the original structure of the $B$-splines and, secondly, obtains computational efficiency both in the basis element evaluations and in the spectral decomposition of a functional signal.

\begin{figure}
    \centering
    \includegraphics[width=0.58\textwidth]{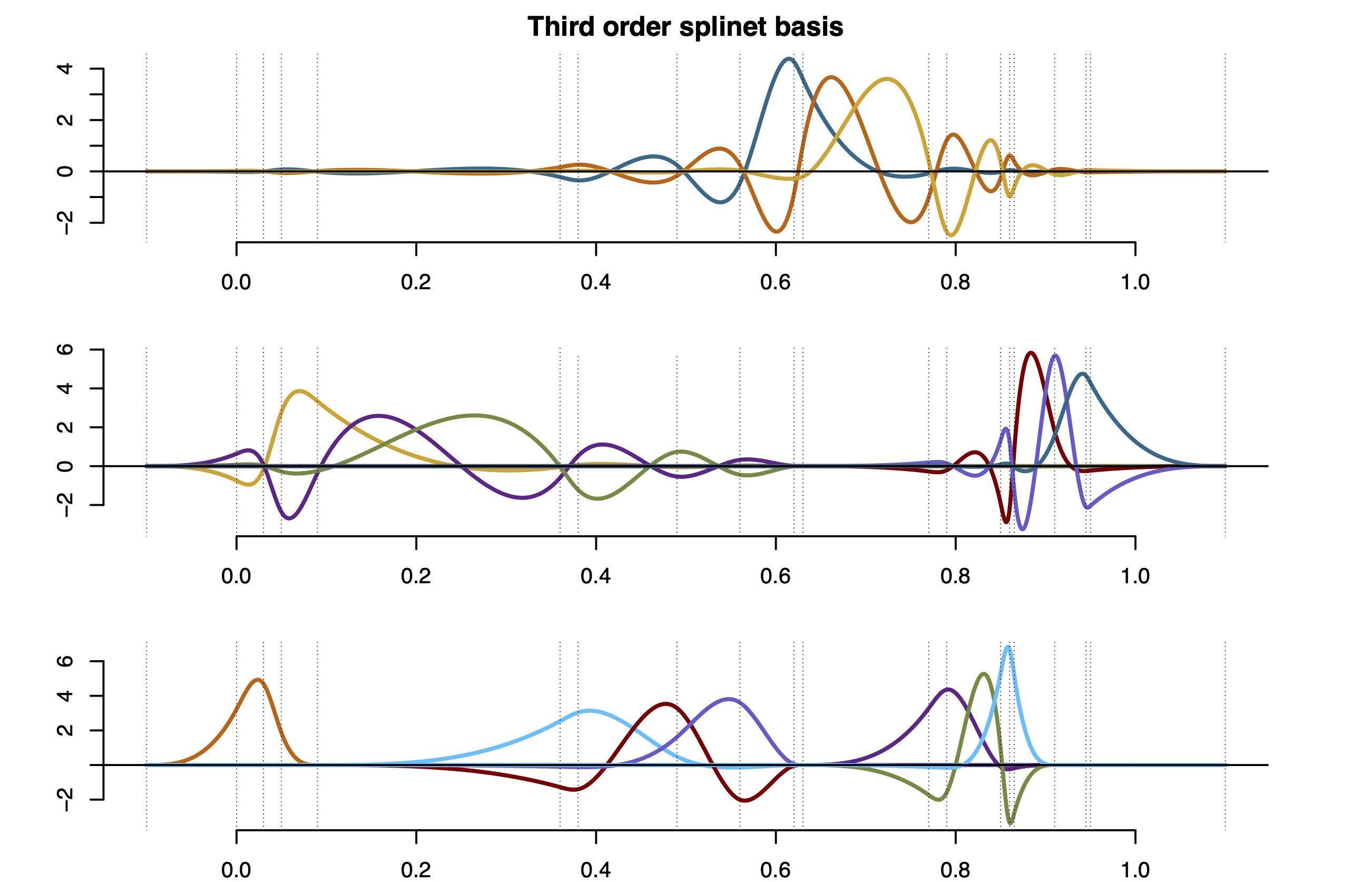}\hspace{-1.8cm}
    \raisebox{-1.2cm}{
    \includegraphics[width=0.51\textwidth]{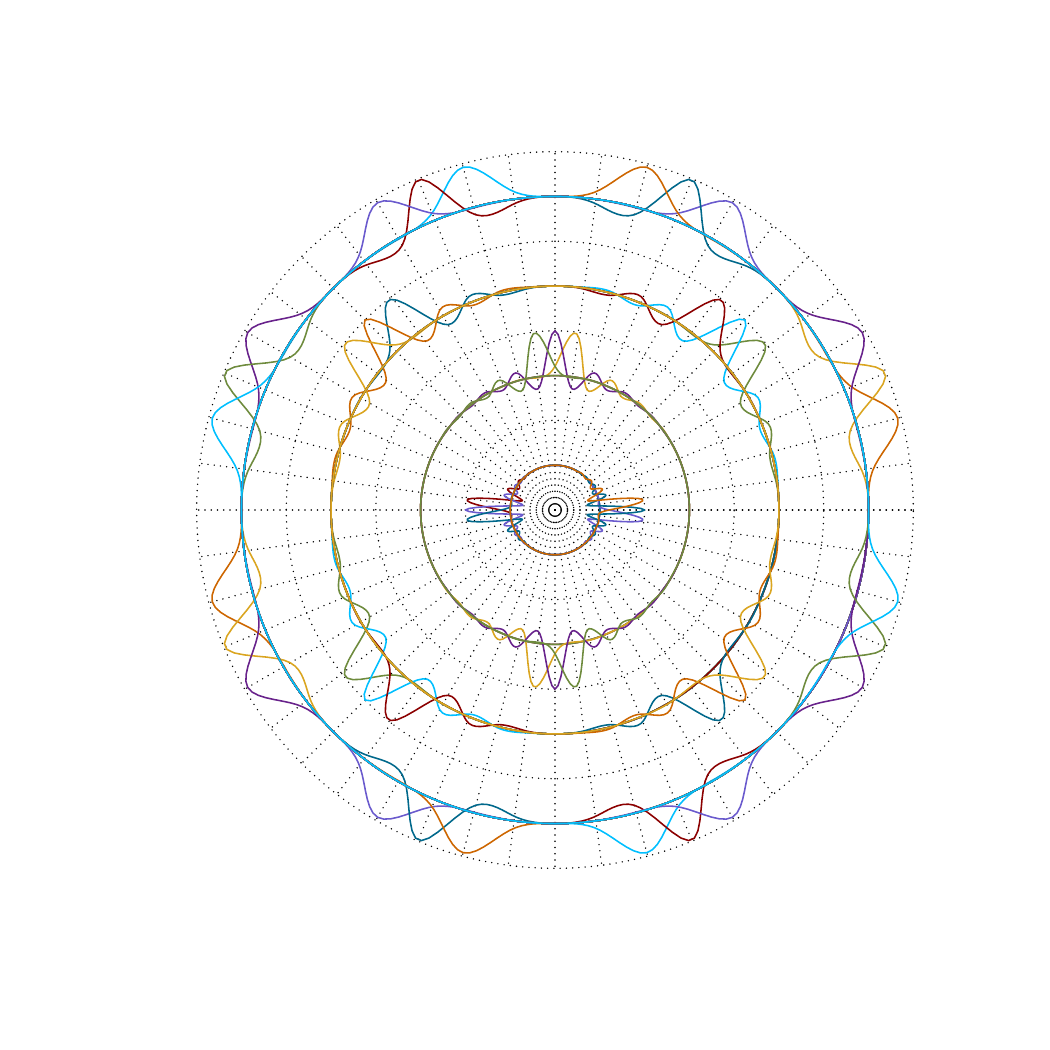}
    } \vspace{-7mm}
    \caption{{\em Splinets} - the orthogonal cubic spline bases presented on dyadic levels: {\it (Left)} A splinet made on irregularly spaced knots on an interval, not a complete dyadic case, see \cite{LIU2022}; {\it (Right)} Periodic splinet on regularly spaced knots presented on circular dyadic levels, a complete dyadic case of 48 O-splines, see also \cite{nassar2023splinets}.}
    \label{fig:splinets}
\end{figure}

Although the method is fundamentally different, the orthogonalization method was inspired by one-sided and two-sided orthogonalization discussed in \cite{mason1993orthogonal}. Since the constructed basis spreads a net of splines rather than a sequence of them, we coin the term {\em splinet} when referring to such a base. It is formally shown that the splinets, similarly to the $B$-splines, feature a locality due to the small size of the
total support. Suppose the number of knots over which the splines of a given order are considered is $n$. In that case, the total support size of the $B$-splines is on the order $O(1)$ with respect to $n$, while the corresponding splinet has the total support size on the order of $\log n$ which is only slightly bigger. On the other hand, the previously discussed orthogonalized bases have the total support size of the order $O(n)$, where $n$ is the number of knots, which is, roughly speaking, also the number of basis functions. Moreover, if one allows for negligible errors in the orthonormalization, then the total support size no longer will depend on $n$, i.e. becomes constant and thus achieves the rate seen for the $B$-splines. An example of a splinet spanned over non-equidistant knots can be seen in Figure~\ref{fig:splinets}~{\it (Left)}.

Beyond splines defined on a subset of real line one can also consider splines defined on a circle, i.e. periodic splines. These are also implemented in the package, as shown in Figure~\ref{fig:splinets}~{\it (Right)}. Nevertheless, the focus of both the package and this presentation is on the splines.
Our goal is to demonstrate that these features of the {\href{https://cran.r-project.org/web/packages/Splinets/index.html}{\tt Splinets}}-package make it an effective tool for functional data analysis (FDA). Not only does it allow for standard methods to be easily implemented, but it also facilitates the exploration of new methodologies that are specific to the choice of splines as functional spaces for analysis.
In fact, the workflow for the classification problem is presented as the main contribution of the paper.
In this workflow, we incorporate the machine learning-based method of the data-driven knot (DDK) selection algorithm that has been proposed in \cite{basna2022data}. 
In the knot selection process, we aim at a data-driven functional space selection that is at the core of our approach and differs from standard approaches that assume some theoretical initial basis selection (for example, Fourier basis, wavelets, or splines with equidistant knots). 
The entire workflow can be summarized in the following steps
\begin{enumerate} [leftmargin=0.4cm]
    \item {\bf Data preparation.} This includes splitting datasets into training and validation/testing parts, and restructuring data representations as needed.
    
    {\it Numerical and computational tools:} Data and case specific, the package {\tt Splinet} is not used.  For example, for images one could utilize the Hilbert curve algorithm implemented in {\tt gghilbertstrings} \textbf{\textsf{R}}-package, that is presented in the additional materials containing the code accompanying the paper.  
    \item {\bf Data-driven knot selection.} This is an effective way of adopting the choice of spline spaces.  We apply the DDK algorithm to select knots across all samples in the training data set within each class and decide on the number of knots. It's important to note that both the number and placement of knots are class-specific. Recognizing the importance of this step is crucial for efficient analysis. However, if efficiency is not a concern and one decides to perform analysis on equally spaced knots this step can be skipped.
    
    {\it Numerical and computational tools:} The main tool here is the \textbf{\textsf{R}}-implementation of the {DDK}-algorithm available  at the GitHub page: \url{https://github.com/ranibasna/ddk}. The {DDK}-algorithm is utilized for stopping the knot selection process.
    \item {\bf Projecting training data into spline spaces.}  After selecting the knots, data can be projected into the spline space spanned by the given knots, specific to the class of a training data point. The order of spline spaces can be chosen depending on the needs and the third order is standard. 
    
    {\it Numerical and computational tools:} Here the package {\tt Splinet} is used to a full extent.
    The main function is {\tt project()} which provides a convenient isometry between the space of splines and the Euclidean space of the same dimension. 
    \item {\bf Classwise functional principal components analysis:} The  FPCA is performed on training data and within each class. Computationally, it is reduced to standard eigenvalue-eigenvector decomposition of positive definite matrices thanks to the isometry between coefficients in the splinets representation of a spline and an Euclidean space of the dimension equal to the size of the splinet. 

    {\it Numerical and computational tools:} Due to the isometry this is done using standard {\tt base} package of \textbf{\textsf{R}}. The central function for this is {\tt eigen()} applied to the sample covariance matrix of the vectors of coefficients of the splinet representation of data points. The package {\tt Splinet} is used to functionally inspect and illustrate the outcomes of the spectral decomposition. 
    \item {\bf Determination of the significant principal components:} 
    After the initial dimension reduction obtained by the choice of knots for the spline representation of the data, the choice of the significant eigenvalues further reduces the dimension of the functional representation of the data. This is done on the validation procedure that uses the accuracy of the classification procedure (described in the next step) on the validation data set. Based on this, the optimal (in the accuracy of classification) number of eigenfunctions is chosen. 

    {\it Numerical and computational tools:} Since this step is performed within the Euclidean space representation of the data, it does not require any special programming tool except for an implementation of the classification procedure that is described next.  The \textbf{\textsf{R}} code that performs all the necessary steps, including the classification procedure, is included in the additional material of this paper.  
    \item {\bf Testing classification procedure:} The classification procedure, already utilized in the previous step, is applied to the testing set following the selection of all hyperparameters, in particular, the number of eigenvectors.
    In this step,  each data point from the testing data is classified by finding the class to which the projection to the space of principal components for each class is closest to the original data point.
    The results are summarized in the confusion matrix and a preliminary assessment of the classification method is given. 

    {\it Numerical and computational tools:} Since the test data points are raw and discrete, all tools used to transform these data into functional data points in the respective spline spaces must be performed. Consequently, most of the computational tools used in the previous steps have to be used again. 
    Visualization and presentation of the confusion matrix, as well as other preliminary evaluations of the classification, can be performed according to the preferences and are not particularly tied to any specific set of tools.
    \item {\bf Final evaluation and conclusions:} This part is dependent on the overall goal of the analysis and is case-specific. 

    {\it Numerical and computational tools:} The tools for performing this post-analysis are case-dependent and goal-specific. Consequently, they cannot be precisely specified and are not limited to concrete packages or routines.  
\end{enumerate}

The organization of the material is as follows. 
We start with a discussion of the functional representation of splines used in the package implementation. 
A discussion of the treatment of support sets for splines with an illustration of efficiency for sparse data is presented. 
We then delve into different spline bases, starting with standard $B$-splines. Following this, we obtain orthonormal bases through the orthogonalization of $B$-splines.
The emphasis is on the orthonormal basis obtained from the B-splines by efficient dyadic algorithms and referred to as a {\tt splinet}. 
The presentation of the important implementation of the projection of functions to the space of splines by means of the developed spline bases follows. The main part of the contribution is presented next, where the workflow for the classification problem is elaborated and applied to a functional data analysis of the fashion MINST data.
Finally, there is also a discussion of the 2D-dimensional features of the data. Some initial steps toward accounting for them are made; see also \cite{basna2023empirically}.
As supplementary material, the commented \textbf{\textsf{R}}-code required to recreate all figures and analyses in the paper is available.
\section{Splines and their representations}
We start with setting the terminology and providing fundamental properties of splines and their representation as implemented in the package {\href{https://cran.r-project.org/web/packages/Splinets/index.html}{\tt Splinets}}.

\begin{figure}
    \hspace{-0.2cm}
    \includegraphics[width=0.375\textwidth]{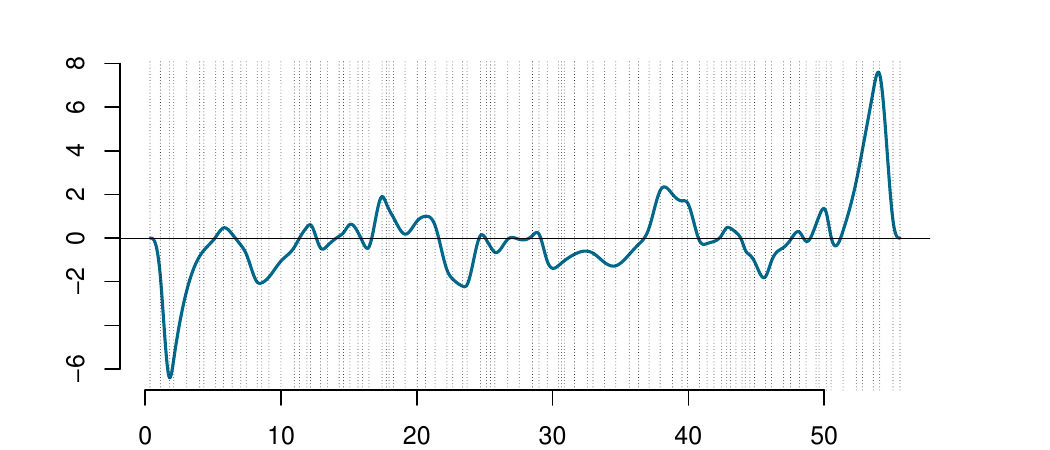}\hspace{-1.05cm}
    \includegraphics[width=0.375\textwidth]{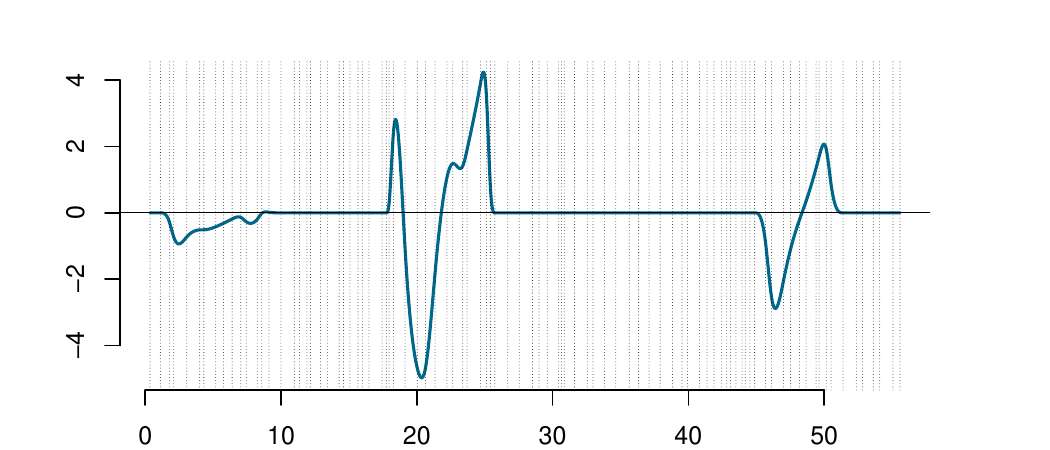}
    \hspace{-1.1cm}
    \includegraphics[width=0.375\textwidth]{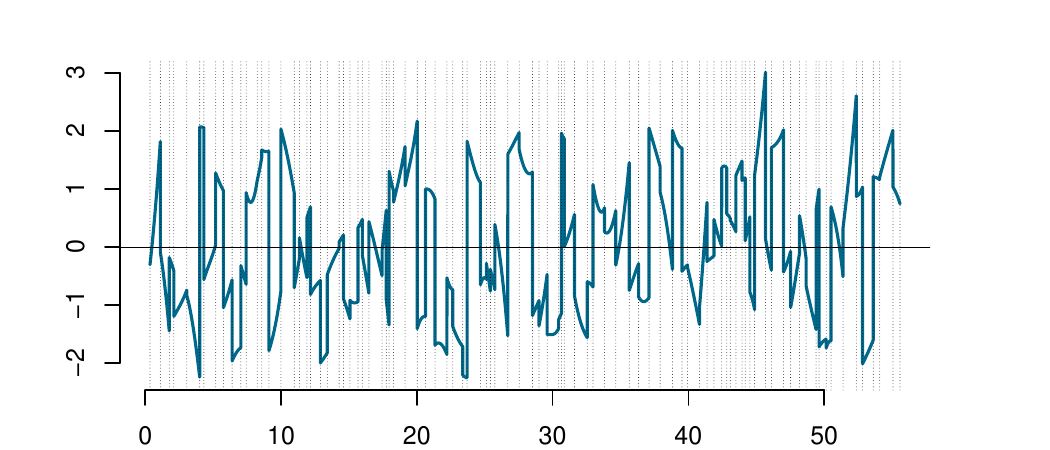}
    \caption{Various functions handled by {\href{https://cran.r-project.org/web/packages/Splinets/index.html}{\tt Splinets}}. {\it Left:} A cubic spline with 90 non-equidistant internal knots; {\it Middle:} A  sparse cubic spline with support over a union of three disjoint intervals; {\it Right:} A piecewise polynomial function that is not a spline.}
    \label{fig:piecwise}
\end{figure}

Splines are piecewise polynomials of a given smoothness order with continuous derivatives, up to this order (exclusive) at the points they interconnect, which are called {\it internal knots}. 
The domain of a spline will be called its {\it range} and is assumed in the paper to be a closed finite interval. 
Additional knots called the  {\it initial knots} and the {\it terminal knots} are located at the beginning and the end of the spline range, respectively.
For a given set of knots, the space of splines is finite-dimensional, with the inner product of the Hilbert space of the square-integrable functions. 
A typical element of a space of splines is presented in Figure~\ref{fig:piecwise} {\it(Left)}.

Although it is beyond the scope of this paper, it is worth mentioning two extensions of functional spaces that are handled by the package. 
The first are splines that do not have full support over the entire range of the considered domain. 
This can be of special utility for the sparse functional data, i.e. the functional data that are nonzero only on a small number of locations.
A good example of such data are mass spectra in proteomics or metabolomics analyses. 
The implementation of the {\tt splinet}-object allows specifying a support of functional spline to be only on a union of disjoint intervals inside of the entire range of domain.
See Figure~\ref{fig:piecwise}~{\it (Middle)} for an example of such a sparse spline. 
The second generalization extends beyond the splines since the {\href{https://cran.r-project.org/web/packages/Splinets/index.html}{\tt Splinets}}-object can be
used to represent any piecewise polynomial function of a given order. All the relevant functions
work properly even if the smoothness at the knots is not preserved, see Figure~\ref{fig:piecwise}~{\it (Right)}. 
In the package various methods of correcting piecewise polynomial functions to make them splines are implemented, including the orthogonal projection to the space of splines.

Due to the continuity requirements, the behavior of a spline between two given knots is necessarily affected by the form of polynomials at the neighboring between-knots subintervals. 
Since the between-knots intervals with terminal knots have only one neighboring between-knots interval, the influence of values over other intervals is not the same.   
To mitigate this biased terminal knots effect, it is natural and, as it will be seen, also mathematically elegant to introduce the zero boundary conditions at the terminal knots for all the derivatives except the highest order one. 
It can be shown that to remove the zero boundary effect, one has to consider splines over the knots obtained through extending by a certain number of knots from both ends of the complete set of knots. 
More specifically, the number of knots that have to be added at each end is equal to the order of the splines. 
Often the knots are added by replicating the terminal knots although there are some serious disadvantages of such an approach.

The most natural and computationally stable way to evaluate the values of a given spline is through the Taylor expansion at the closest knot. 
For this, it is convenient to have all derivatives at a knot directly accessible. 
To achieve this, the {\href{https://cran.r-project.org/web/packages/Splinets/index.html}{\tt Splinets}} implementation of a functional spline uses an object that holds the matrix of the derivatives at the knots. 
Such a matrix needs to satisfy certain conditions to ensure that the derivatives at the knots are continuous. 
In this section, we review the mathematical properties of splines that are fundamental for implementation.

\subsection{Splines with zero-boundary conditions at the endpoints}
The splines involve knots at which the polynomials smoothly connect. 
A set of such knots is represented as a vector $\boldsymbol \xi$ of $n+2$ ordered values. 
As already mentioned, there are two alternatives, but in a certain sense equivalent, requirements on the behavior of a spline at the endpoints of its range.
In the first, no boundary conditions are imposed. The main problem in this unrestricted approach is that the polynomials at both ends of the considered range do not `sense' the same restrictions from the neighbors as the polynomial residing further from both endpoints. This is due to the fact that at the endpoints the `neighbors'  are only present from one side.
Another approach, favored in this work, is by putting zero boundary restrictions on the derivatives at the endpoints.
This approach is mathematically equivalent to the first one in a limiting sense, when the $k$ initial knots and the $k$ terminal knots converge to the beginning and the end of the range, respectively.  Moreover and most importantly, the approach is structurally elegant and thus easier to implement.
For all these reasons, it is used in our package.   However, the main object of the package which is {\tt splinets}

Throughout the rest of the paper, we impose on a spline and all its derivatives of the order smaller than the spline smoothness order the value of zero at both the endpoints of the range. 
In this case, if we consider knots $\boldsymbol \xi=(\xi_{0},\dots, \xi_{n+1})$
it is important to assume that  $n\ge k$, in order to have enough knots to define at least one non-zero spline with the $2k$ zero-boundary conditions at the endpoints.
Indeed, if $n=k-1$, then we have $k+1$-knots yielding $k$ between knot intervals. 
On each such interval, a spline is equal to a polynomial of order $k$.
The dimension of the space of such piecewise polynomial functions is $k(k+1)$. However, at each internal knot, there are $k$ equations to make derivative up to order $k$ (excluding) continuous. 
This reduces the initial unrestricted polynomials dimension by $(k-1)k$ dimensions to $2k$, but there are $2k$ equations for the derivatives to be zero at the endpoints.
We conclude that the dimension of the spline space is eventually reduced to zero, meaning that there is only a function trivially equal to zero in this space.
From now on $\mathcal S^{\boldsymbol \xi}_{ k}$ stands for the $n-k+1$ dimensional space of the $k$-smoothed splines with the zero boundary conditions at the terminal knots of in the ordered knots given in vector $\boldsymbol \xi$.
Whenever showing the dependence on either $k$ or $\boldsymbol \xi$ or both is not important, they will be dropped from the notation. Thus, for example, $\mathcal S$ stands for $\mathcal S^{\boldsymbol \xi}_{ k}$ if both $k$ and $\boldsymbol \xi$ are clear from the context. 

The splines can be represented in a variety of ways. In \cite{Qin}, a general matrix representation was proposed that allows for efficient numerical processing of the operations on the splines. 
This was utilized in \citet{Zhou} and  \citet{Redd} to represent and orthogonalize $B$-splines that were implemented in the \textbf{\textsf{R}}-package \href{https://CRAN.R-project.org/package=orthogonalsplinebasis}{\it Orthogonal B-Spline Basis Functions}. 
In our approach, we propose to represent a spline in a different way.
Namely, we focus on the values of the derivatives at knots and the support of a spline.  
The goal is to achieve better numerical stability as well as to utilize the discussed efficiency of base splines having support only on a small portion of the considered domain.

The fundamental fact that we use here is that for a given order, say $k$, and a vector of knot points $\boldsymbol \xi=\left(\xi_0,\dots, \xi_{n+1}\right)$, a spline $S\in \mathcal S$ is uniquely defined by the values of the first $0,\dots, k$ derivatives at the knots.
Here, by a natural convention that we use across the paper, the $0$ derivative is the function itself. 
The values of derivatives at the knots allow for the Taylor expansions at the knots but they cannot be taken arbitrarily due to the smoothness at the knots.

The values of the derivatives at the knots are at the center of the Taylor expansion representation of a spline.  
Therefore the matrix of the derivative values has become the main component of an object belonging to our main class in the package.
Let us assume that we consider a spline with the full support range, i.e. not vanishing on any subinterval of the entire range of knots. 
For any spline function $S\in \mathcal S_k^{\boldsymbol \xi}$, we consider $\mathbf s_j=(s_{0j},\dots, s_{n+1j})$ is an $n+2$-dimensional  vector (column) of values of the $j^{\rm th}$-derivative of $S$, $j=0,\dots, k$, at the knots given in vector $\boldsymbol \xi= \left(\xi_0,\dots, \xi_{n+1}\right)$ that, as a general convention for all vectors (also the convention in \textbf{\textsf{R}},  is treated as a $(n+2)\times 1$ column matrix. 
These columns are kept in a $(n+2)\times (k+1)$ matrix 
\begin{equation}
\label{eq:spmat}
\mathbf S\stackrel{def}{=}\left [ \mathbf s_0 \mathbf s_1 \dots  \mathbf s_k \right].
\end{equation} 

More specifically, the main object {\tt Splinets} implemented through the S4 system for the OOP in \textbf{\textsf{R}} is defined through {\tt setClass} function with the following slots
{\footnotesize \begin{verbatim}
representation(knots="vector", smorder="numeric", equid="logical", type = "character",
           supp="list", der="list", taylor = "matrix", type = "character", epsilon="numeric"),
         \end{verbatim} }
\vspace{-.4cm}\noindent which represents a collection of splines,  all built over the same knots given in {\tt knots}, of the smoothness order $k$ given in {\tt smorder}.
Further {\tt supp}  is the list of matrices having row-wise pairs of the endpoints of the intervals, the union of which constitutes the support set of a particular spline,  and the flag {\tt equid} informs about the equally placed knots, for which the computation can be significantly accelerated. 
The matrices of the derivatives at the knots inside the support sets are given in the list {\tt der} of 
 matrices, where an element in the list refers to a particular spline in our collection and the length of the list corresponds to the number of splines in the object.  
Descriptions of other fields are given in the {\href{https://cran.r-project.org/web/packages/Splinets/index.html}{\tt Splinets}}-package but are not crucial for this presentation.

Consequently, the above object corresponds to a spline function $S$ of order $k$ over knots in $\boldsymbol \xi=(\xi_0,\dots, \xi_{n+1})$ that is identified as 
$$
S=\left\{k, \boldsymbol \xi,\mathcal I, \mathbf s_0,\mathbf s_1, \dots, \mathbf s_k\right\},
$$
where $\mathcal I=\{(i_1, i_1+m_1+1), \dots,(i_ N,i_N+m_N+1) \}$ is a sequence of ordered pairs of indexes in $\{1,\dots,n+2\}$ representing the intervals, the union of which is the support of a spline, i.e. the minimial closed set outside  of which spline vanishes. 

\subsection{Splinets -- orthonormal bases of splines}
The direct approach to building splines requires a lot of care and often can be cumbersome. 
A more efficient approach to building splines is through functional bases of splines. 
There are many possible choices of such bases but the most popular are the $B$-splines. 
Despite having many advantages, the $B$-splines do not constitute an orthogonal basis. 
Our main contribution is to implement an optimal orthogonalization of the $B$-splines introduced in \cite{LIU2022}.
The presentation of this spline basis benefits from organizing them in the form of a net. 
In this framework, the derived orthogonal bases of splines are referred to as the splinets. 

The most convenient way to define the $B$-splines on the knots $\boldsymbol \xi$,  is through the splines with the boundary conditions and using the recurrence on the order. 
Namely, once the $B$-splines of a certain order are defined, then the $B$-splines of the next order are easily expressed by their `less one' order counterparts. 
In the process, the number of the splines decreases by one, and the number of the initial conditions (derivatives equal to zero) increases by one at each endpoint. 
Let $B^{\boldsymbol \xi}_{l,k}$, is the $l$th $B$-spline of the order $k$, $l=0,\dots ,n-k$. 
For the zero-order splines, the $B$-spline basis is made of indicator functions 
\begin{equation}
\label{eq:indi}
B^{\boldsymbol \xi}_{l,0}=\mathbb I_{(\xi_{l},\xi_{l+1}]}, ~~~l=0,\dots n,
\end{equation}
 for the total of $n+1$-elements and zero initial conditions. 
Clearly, the space of zero order splines (piecewise constant functions) is $n+1$ dimensional so the so-defined zero order $B$-splines constitute an orthogonal basis.

The following recursion relation leads to the $B$-splines of an arbitrary order $k \le n$.
Suppose now that we have defined $B^{\boldsymbol \xi}_{l,k-1}$, $l=0,\dots,n-k+1$. 
The $B$-splines of order $k$ are defined, for $l=0,\dots,n-k$, by
\begin{equation}
\label{eq:recspline}
B_{l,k}^{\boldsymbol \xi }(x)
=
 \frac{ x- {\xi_{l}}
  }{
  {\xi_{l+k}}-{\xi_{l}} 
  } 
 B_{l,k-1}^{\boldsymbol \xi}(x)+
  \frac{{\xi_{l+1+k}}-x}{
 {\xi_{l+1+k}}-{\xi_{l+1}} 
 } 
 B_{l+1,k-1}^{\boldsymbol \xi}(x).
\end{equation}

It is also important to notice that the above evaluations need to be performed only over the joint support of the splines involved in the recurrence relation.
The recurrent structure of the support is as follows. 
For zero order splines, the support of $B_{l,0}^{\boldsymbol \xi}$ is clearly $[\xi_l,\xi_{l+1}]$, $l=0,\dots, n$. 
If the supports of $B_{l,k-1}^{\boldsymbol \xi}$'s are $[\xi_l,\xi_{l+k}]$, $l=0,\dots,n-k-1$, then the support of $B_{l,k}^{\boldsymbol \xi}$ is the joint support of $B_{l,k-1}^{\boldsymbol \xi}$ and $B_{l+1,k-1}^{\boldsymbol \xi}$, which is $[\xi_l,\xi_{l+1+k}]$,  $l=0,\dots,n-k$.
In order to translate these recursive relations to the relations between the matrices of the derivatives at the knots, we need the following result on the derivatives of the $B$-splines. 

\begin{proposition}
\label{prop:dersp}
For $i= 0,\dots, k$ and  $l=0,\dots,n-k$:
\begin{multline}
\label{eq:recder}
\frac{d^iB_{l,k}^{\boldsymbol \xi}}{dx^i}(x)
=
\frac{i}{\xi_{l+k}-\xi_l}\frac{d^{i-1}B_{l,k-1}^{\boldsymbol \xi}}{dx^{i-1}}(x)+\frac{ x- {\xi_{l}}
  }{
  {\xi_{l+k}}-{\xi_{l}} 
  } 
\frac{d^{i}B_{l,k-1}^{\boldsymbol \xi}}{dx^{i}}(x)
+\\
+
\frac{i}{\xi_{l+1}-\xi_{l+k+1}}\frac{d^{i-1}B_{l+1,k-1}^{\boldsymbol \xi}}{dx^{i-1}}(x)+\frac{ {\xi_{l+k+1}-x}
  }{
  {\xi_{l+k+1}}-{\xi_{l+1}} 
  } 
\frac{d^{i}B_{l+1,k-1}^{\boldsymbol \xi}}{dx^{i}}(x).
\end{multline}
The support of ${d^iB_{l,k}^{\boldsymbol \xi}}/{dx^i}$ is $[\xi_{l},\xi_{l+k+1}]$ and
if $i=k$, then $d^{i}B_{l+1,k-1}^{\boldsymbol \xi}/dx^{i}\equiv 0$.
\end{proposition}


\begin{figure}[t!]
\begin{center}
\includegraphics[width=0.495\textwidth, height=4.5cm]{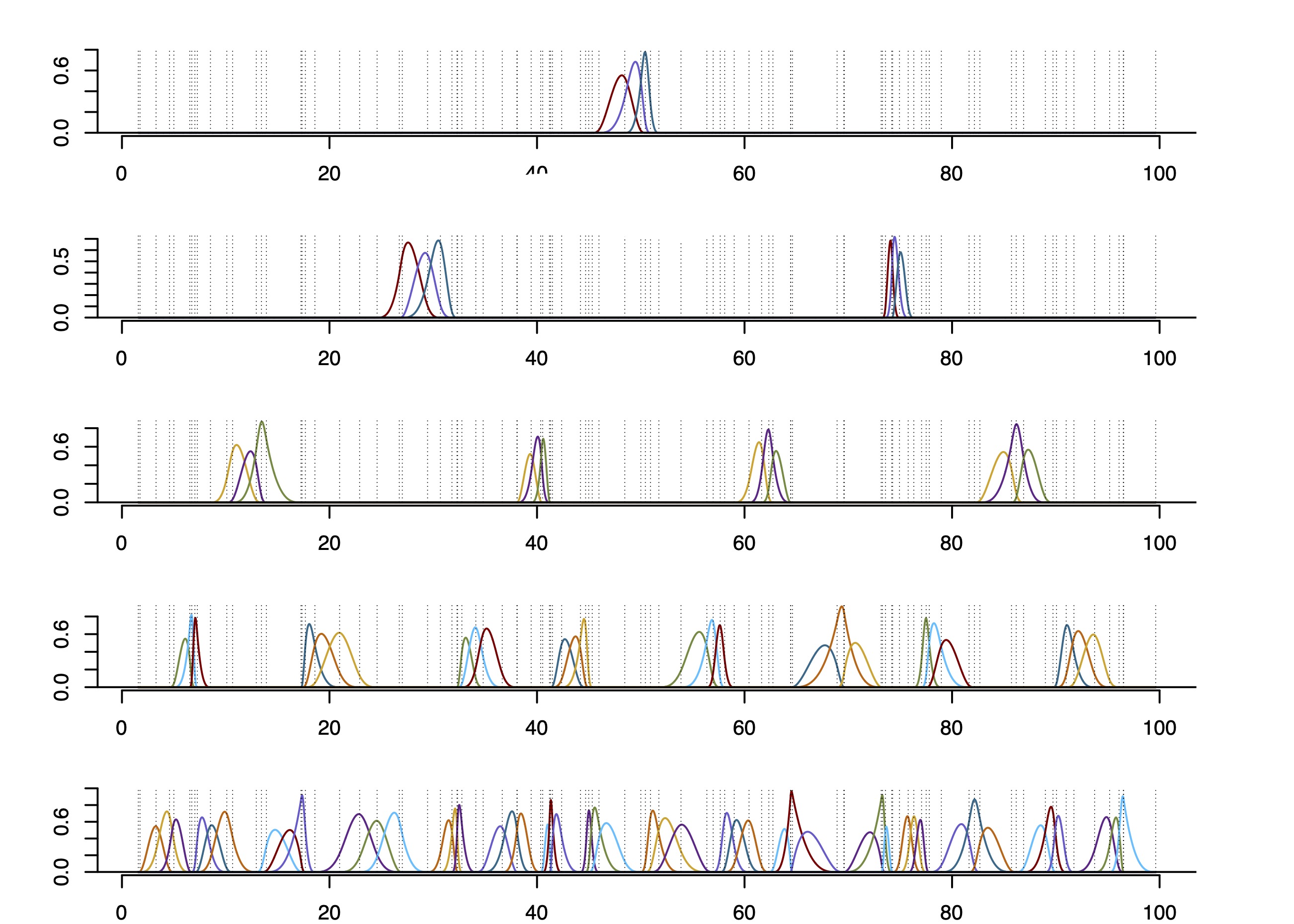} 
\includegraphics[width=0.495\textwidth, height=4.5cm]{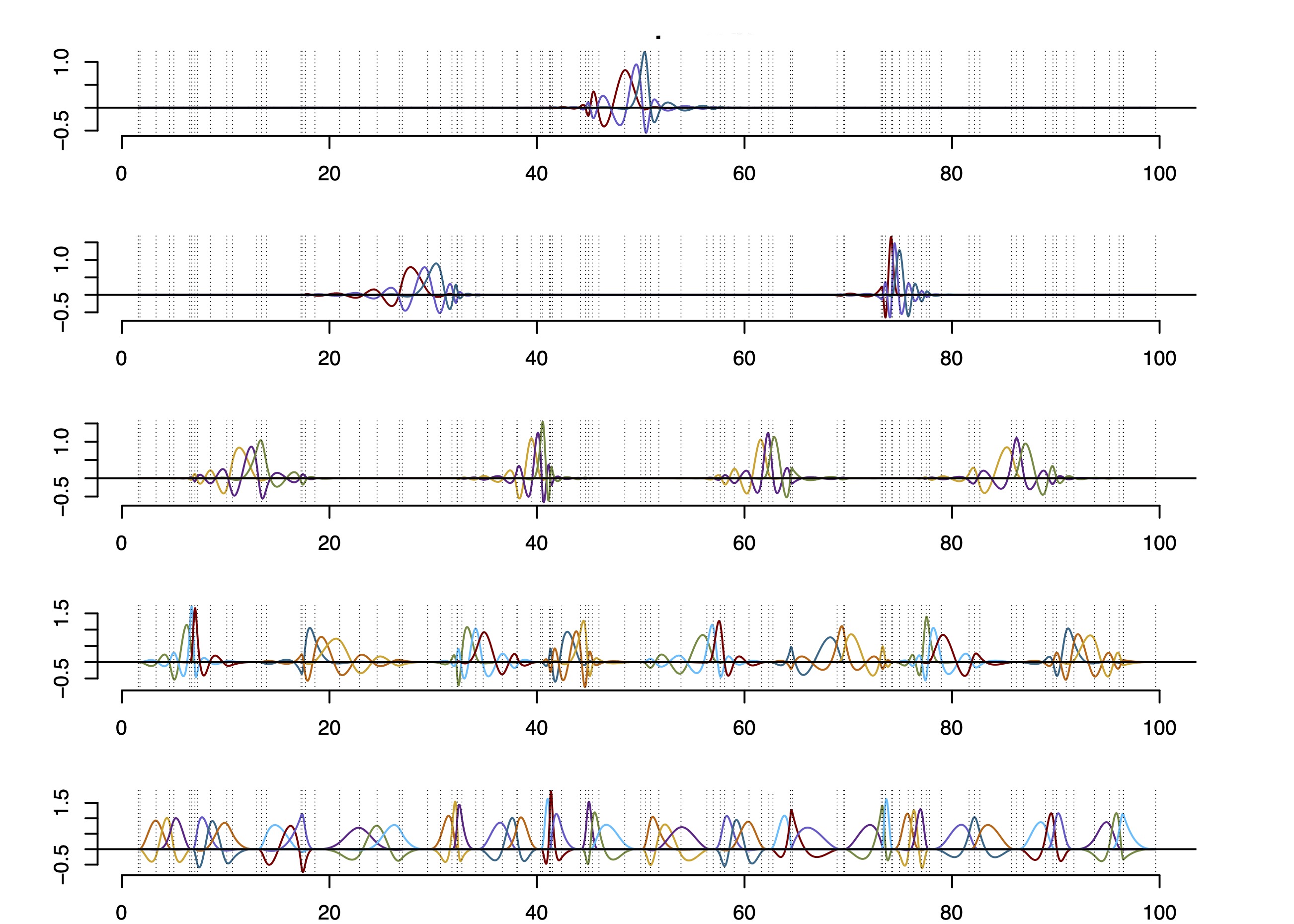} 
\end{center}   
    \caption{Cubic spline bases presented graphically on a dyadic net. The case of $n=k2^N-1=95$, $k=3$, $N=5$ which is the number of layers in the dyadic structure seen in the figures. {\it Left:} The $B$-spline basis; 
  {\it Right:} The corresponding splinet.}
  \label{fig:Orth}
\end{figure}


Consider the zero$^{\rm th}$ and first order $B$-splines with the boundary conditions  $B_{0,l}^{\boldsymbol \xi}$,  $B_{1,r}^{\boldsymbol \xi}$, where $l=0,\dots, n$, $r=0,\dots, n-1$. 
Then the corresponding $(n+2)\times 1$ and $(n+2)\times 2$ matrices are
\begin{equation*}
\mathbf S^{(0,l)}=\begin{bmatrix} 0 \\ \vdots \\ 0 \\ 1 \makebox[0in]{~~~\hspace{14mm}$\leftarrow l+1$} \\ 0 \\ \vdots\\  0
\end{bmatrix} \hspace{1.2cm} ,~~l\le n, \hspace{3mm} 
\mathbf S^{(1,r)}=\begin{bmatrix} 
0 & 0  \\
 \vdots & \vdots \\
  0 &  \frac{1}{\xi_{r+1}-\xi_{r}}\\ 
  1 & \frac{-1}{\xi_{r+2}-\xi_{r+1}} \makebox[0in]{~~~\hspace{28mm}$\leftarrow r+2$, $r < n$.} \\ 
  0  &  0\\
   \vdots & 
  \vdots \\  0 & 0
\end{bmatrix}
\end{equation*} 

Using the recurrent relation \eqref{eq:recder} between the derivatives of $B$-splines, one can generalize the recurrence between matrix representation of the $B$-splines to the arbitrary order of splines. 
Using one sided representation of the splines, we have 
\begin{multline*}
\label{eq:matrec}
\mathbf S_{\cdot j}^{(k,l)}=\frac{1}{\xi_{l+k}-\xi_l}\left({j}\cdot \mathbf S_{\cdot j-1}^{(k-1,l)} + \boldsymbol \Lambda_l  \mathbf S_{\cdot j}^{(k-1,l)} \right)
+
\frac{1}{\xi_{l+1}-\xi_{l+k+1}}\left({j}\cdot \mathbf S_{\cdot j-1}^{(k-1,l+1)} + \boldsymbol \Lambda_{l+k+1}  \mathbf S_{\cdot j}^{(k-1,l+1)} \right),
\end{multline*} 
where $l=0,\dots, n-k$, $j=0,\dots , k$ and the diagonal $(n+1)\times (n+1)$ matrices $\boldsymbol \Lambda_l$'s have $(\xi_0-\xi_l , \dots , \xi_n-\xi_l)$ on the diagonal.
Here we assume that if $j=k$, then $\mathbf S_{\cdot j}^{(k-1,l)}$ is a column made of zeros as the $k^{\rm th}$ derivatives of the $(k-1)^{\rm th}$ order spline is always zero. The algebraic relation above is implemented in the package to construct the $B$-spline basis of order {\tt k}. 
One uses 
\vspace{-.2cm}
\begin{verbatim}
so = splinet(xi, k); Bsplines=so$bs
\end{verbatim} \vspace{-.2cm}
 where {\tt splinet()} generates  both the $B$-spline basis and its orthognalization, and organizes them as a list. 
The element in the list labeled $\tt bs$ is always the {\tt Splinets}-object corresponding to the $B$-spline basis.
The visualization of this basis with $93$ $B$-splinets is shown in Figure~\ref{fig:Orth} {\it (Left)}.

The orthonormalized bases implemented in this package are obtained by one of the following three orthogonalization procedures applied to $B$-splines. 
The first one is simply the Gram-Schmidt orthogonalization performed on the $B$-splines ordered by their locations, the second one is a symmetric (with respect to the knot locations) version of the Gram-Schmidt, and, finally, the dyadic orthogonalization into a {\em splinet} which is our preferred method. 
We will not discuss the first two orthonormalization methods as they have been included in the package mostly because of historical reasons. 
In the object representation of collections of splines, i.e. in the {\tt Splinets}-class, the field {\tt type} specifies which of the orthonormal basis one deals with. 
The function  {\tt splinet()} is generating the proper basis with the default form 
\vspace{-.2cm}
\begin{verbatim}
so=splinet(xi); Bsplines=so$bs; Splinet=so$os
\end{verbatim} \vspace{-.2cm}
and returning a list of two {\tt Splinets} objects, {\tt so\$bs} and {\tt so\$spnt} build over the ordered knots {\tt xi}.
The first object represents the basis of the standard cubic $B$-splines and is thus not orthogonal. 
The second one represents the recommended orthonormal basis, which is referred to as a cubic splinet. 
In Figure~\ref{fig:Orth}~{\it (Right)}, one can see the splinet obtained from the $B$-splines given in the left-hand side.

The algorithm is most naturally described for the dyadic structure of knots that assumes that for some positive integer $N$ we have $k2^N-1 = n$, where $k$ is the order of the splines and $n$ is the number of the internal knots, i.e. we do not count the endpoints.
The resulting $OB$-splines are then located on a dyadic net with the $N$ levels featuring increasing support set sizes and a decreasing number of basis elements.  
The $OB$-splines are also grouped into $k$-tuples of the neighboring splines. 
All these features are best seen in Figure~\ref{fig:Orth}, {\it (Right)}. 
We observe that each of the $OB$-spline in the splinet inherits its location from the corresponding $B$-spline.
The location is naturally represented by the middle knot in their support if the number of the knots in the support is odd, or by the average of the two middle knots if that number is even. 
These locations can be used to present any spline basis on a dyadic-net graph. 
In fact, the algorithms are implemented in such a way that any set of knots leads to a splinet that can be represented on a dyadic net which may be incomplete if the number of knots does not satisfy the dyadic case restriction. 
 
 The special case of equally spaced knots corresponds to the Gramian (matrix of the inner products) for the $B$-splines being a Toeplitz matrix.
 In this case, certain parts of the algorithms can be significantly accelerated since the evaluation has to be performed only for one $k$-tuple instead of $2^{N-l}-1$ of them at each support level $l$ in the dyadic net. 


\subsection{Projection to space of splines}
The splines with a given knot selection constitute finite dimensional spaces of the Hilbert space of all square integrable functions. 
Thus, any such function can be projected in an orthogonal fashion into the linear space of splines spanned over a particular set of knots. 
Functional data analysis typically begins with projecting the discretized data to a  finite dimensional functional subspace -- the projection becomes a fundamental operation for carrying out statistical data analysis. 
One can also perform a projection to smooth the data and there is a plethora of methods to target this goal. 
In the package, we have implemented orthogonal projection in the function {\tt project()}. 
Since actual functional data can be represented in a variety of ways, the projection itself depends on the input format and, more specifically, the way the inner product of the input with a spline is numerically evaluated.  
The input of the {\tt project()} function can be either {\tt Splinets}-objects or columns of pairs representing arguments and values of a discretization of functional data.

Independently of the input, the output of the function {\tt project()} is a list, say {\tt projsp}, made of the three components:
\begin{description}
\item{\tt projsp\$coeff} -- the matrix of coefficients of the decomposition in the selected basis,
\item{\tt projsp\$basis} -- the {\tt Splinets}-object representing the selected basis,
\item{\tt projsp\$sp} -- the {\tt Splinets}-object representing the projection of the input in the projection spline space.
\end{description}
Additional details, including the knots, the order, and the type of the basis, can be readily obtained from the second component.
Many of the algebraic operations on the splines are more conveniently performed on the matrix of coefficients of their spline basis representations, rather than directly on the {\tt Splinets}-objects. 
The coefficient matrix {\tt projsp\$coeff} can be utilized for such computations and the corresponding linear combination of {\tt projsp\$basis} can be used whenever the functional form of the result is needed. 

\noindent{\it Embedding a spline into higher dimension spaces of splines}
The first function that we discuss is an embedding rather than a projection. 
One of the most interesting aspects of the spline spaces is their agility following different choices of the knots.
Yet, when employing splines for functional data analysis, the focus is typically on splines with a fixed, often equally spaced, set of knots, rather than exploring this adaptability.
In the proposed package, we provide tools to thoroughly explore the properties of splines under various knot choices. 
 Any spline remains a spline of the same order when considered on a set of knots larger than the original.
However, this changes {\tt Splinets} representation of the so-refined spline. 
It is thus important to have a function that embeds a given spline into the bigger space of splines residing on a refined set of knots. 
In the package, the function {\tt refine()} allows splines from smaller spaces to be conveniently represented in larger, more refined spaces.

\noindent{\it Basis decomposition.} The simplest projection obtained through {\tt project()} is not, strictly speaking, a projection but rather a decomposition of a {\tt Splinets}-object to coefficients in the given basis. 
If {\tt sp} is a {\tt Splinets}-object, then 
\vspace{-.2cm}
\begin{verbatim}
bdsp=project(sp); bdsp2=project(sp,type='bs'); 
bdsp3=project(sp,type='gsob'); bdsp4=project(sp,type='twob'); 
\end{verbatim} \vspace{-.2cm}
have as its main output the matrices of coefficients $a_{ji}$,  such that the $j^{\rm th}$ input-spline {\tt sp} has the form
$$
\sum_{i=1}^{n-k+1} a_{ji} OB_{i},
$$
where $j$ indexes the input splines, $n$ is the number of the internal knots, $k$ is the smoothness order, and $OB_{i}$ is the selected basis of splines controlled by the input {\tt type}. Possible choices of bases are the splinet (default), the $B$-splines, the one- (Gram-Schmidt), and two-sided orthonormal bases. All of these are built on the same knots as the input spline.

\begin{figure}[t!]
\raisebox{-4mm}{  \includegraphics[width=0.34\textwidth]{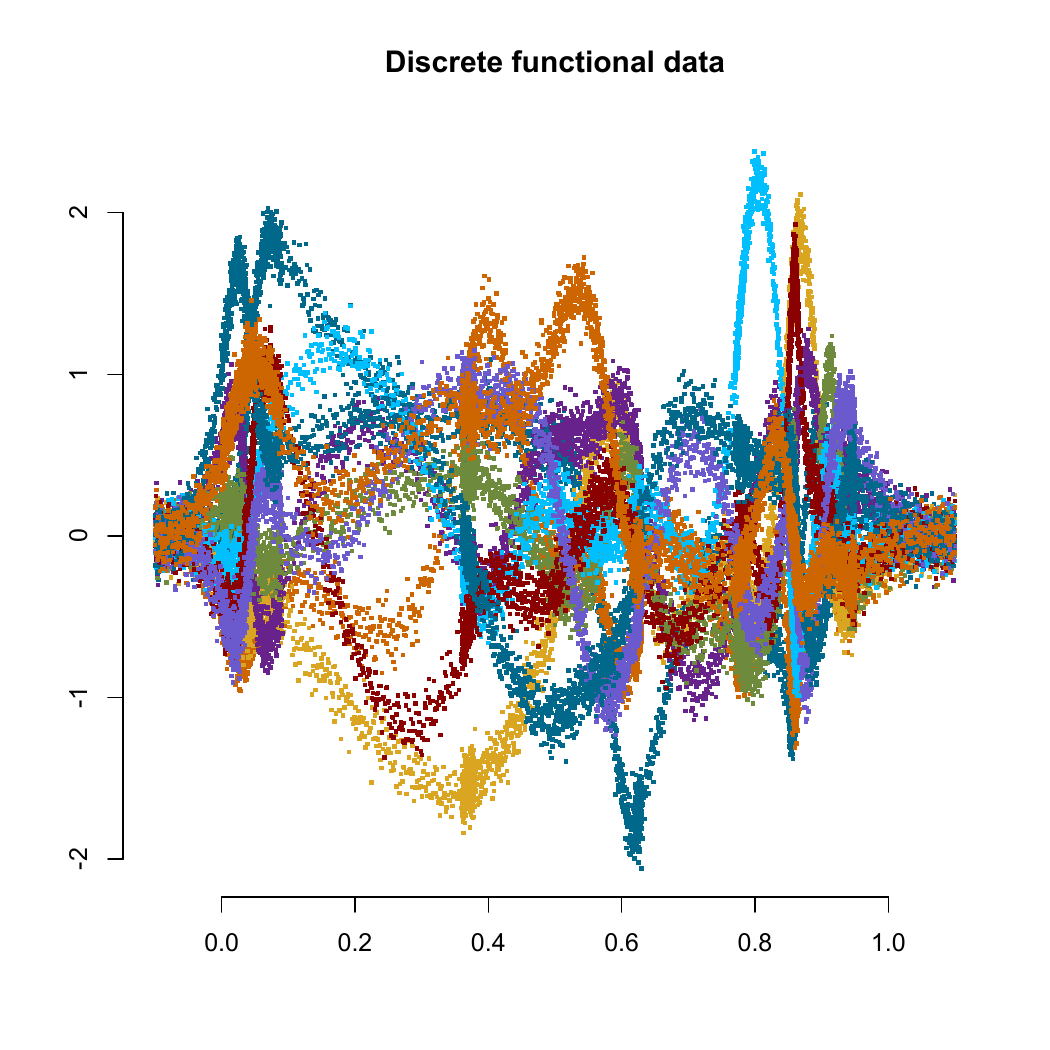} } 
\hspace{-6mm} \includegraphics[width=0.35\textwidth,height=0.3\textwidth]{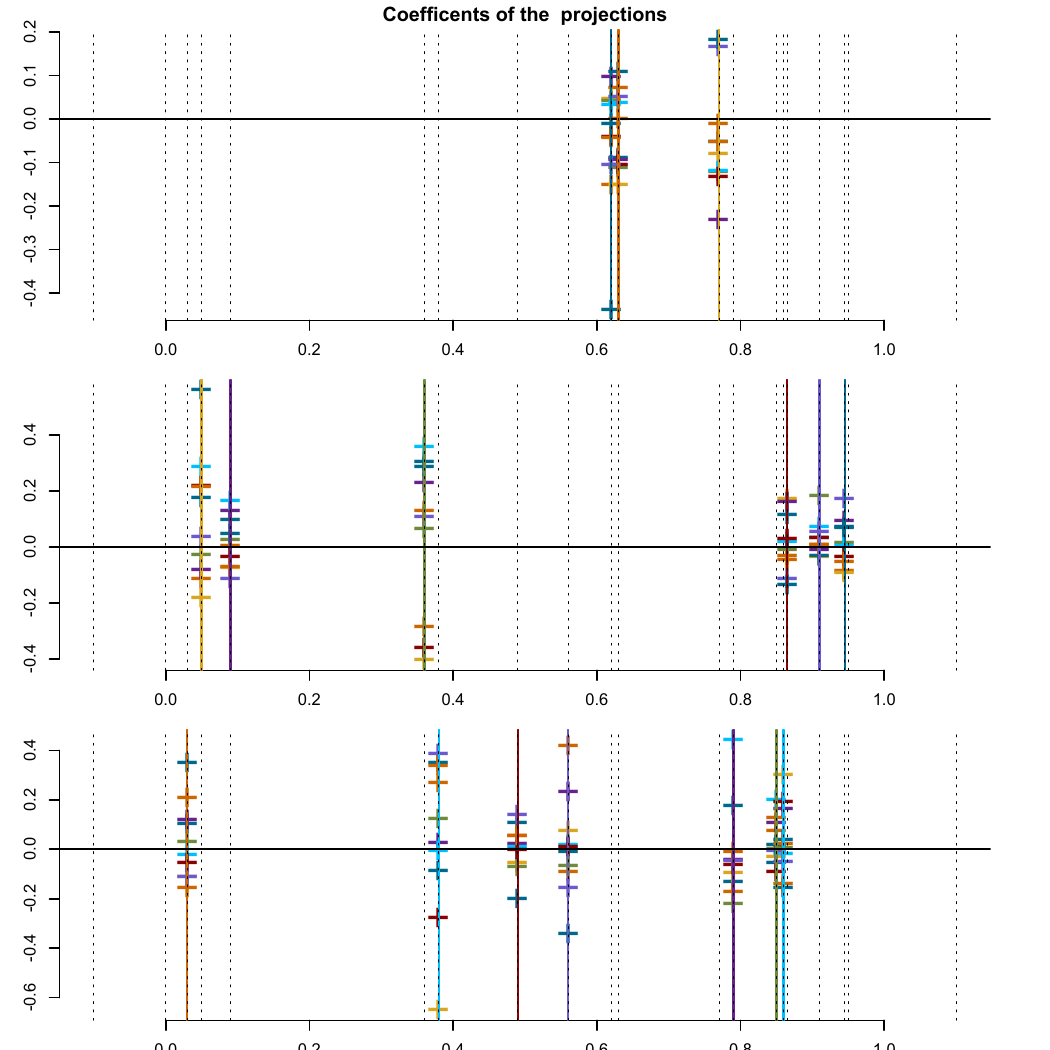} \hspace{-7mm} 
\raisebox{-0mm}{     \includegraphics[width=0.4\textwidth,height=0.3\textwidth]{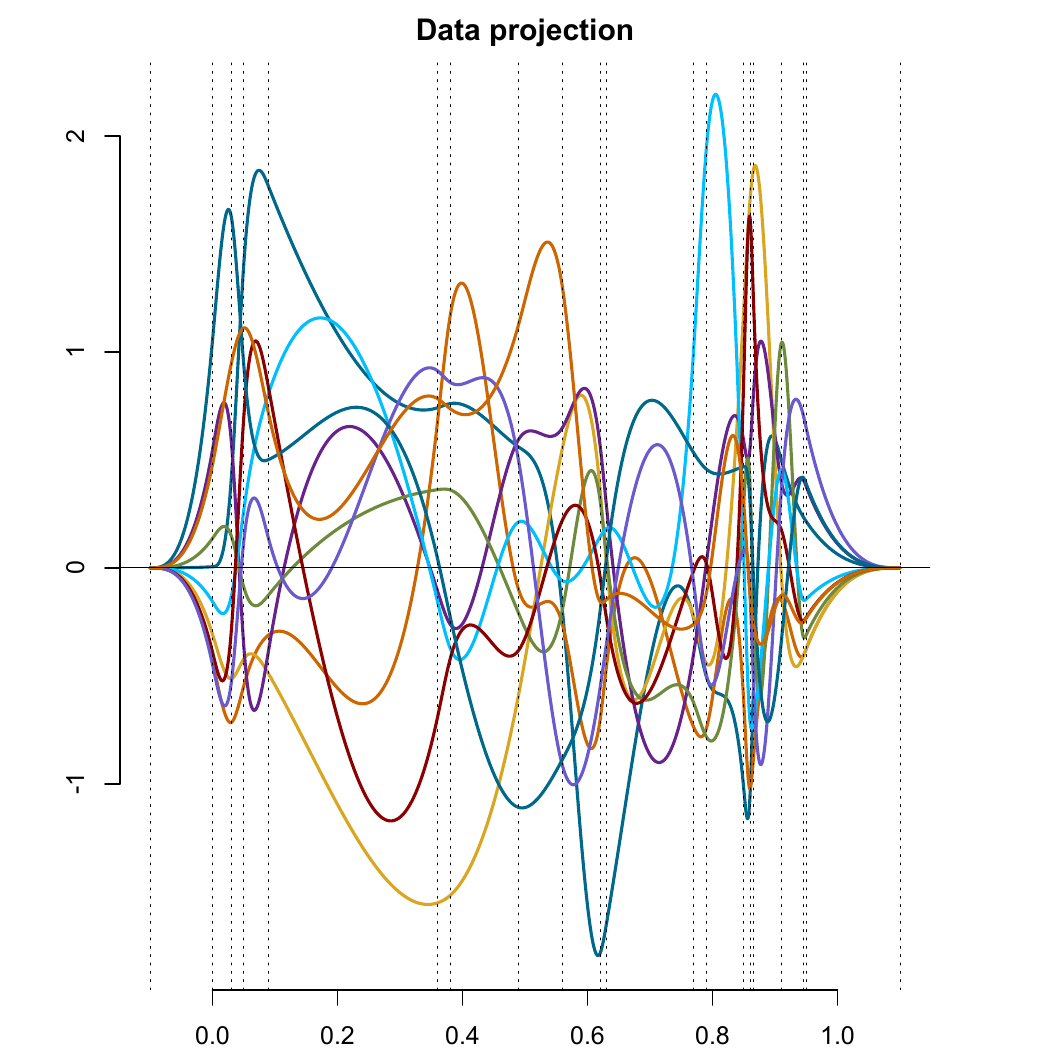} }
\vspace{-0.35cm}
    \caption{ The graphical output of the {\tt project()} function applied to discretized data projected to the spline space spanned on the splinet seen in Figure~\ref{fig:splinets}~{\it (Left)}. 
    {\it Left:}\, The original data $S_j$, $j=1,\dots, 10$. {\it Middle:} The projection coefficients $(a_{ji})_{j,i=1}^{10,n-k+1}$ for the data corresponding to the basis elements in the splinet.  
    {\it Right}\, The projections. 
    The vertical lines represent the knot locations in the projection space. 
    }
    \label{fig:Projectiion}
\end{figure}

\noindent{\it Projecting splines.} 
The projection of {\tt Splinets}-objects over a given set of knots to the space of splines over a different set of knots is obtained as the orthogonal projection.
Namely, if $S$ is  the input {\tt Splinets}-object then the output is denoted as $\mathbf P S$ where $\mathbf P$ is the orthogonal projection to the space spanned by the spline space build by the second set of knots
\begin{equation}\label{projection}
\mathbf P S =\sum_{i=1}^{n-k+1} a_{ji} OB_{i}, \,\, (S-\mathbf P S) \bot \,\mathbf P S.
\end{equation}
This is an extension of the previous case since the output functions may belong to a different space than the input functions.  
The result is obtained by embedding both the input splines and the projection space to the space of splines that contains both and evaluating the inner products between functions in this space. 
The space of splines that contains both is built over the union of the two sets of knots and uses the package function {\tt refine()} for that purpose.
 The following code will lead to the result if {\tt knots} are different from {\tt sp@knots}
 \vspace{-.2cm}
\begin{verbatim}
bdsp=project(sp,knots); bdsp2=project(sp,knots,type='bs'); 
bdsp3=project(sp,knots,type='gsob'); bdsp4=project(sp,knots,type='twob'); 
\end{verbatim}\vspace{-.2cm}
The results are represented in the spline space build over {\tt knots} and thus the {\tt Splinets}-object in the output representing the projection spline satisfies {\tt bdsp\$bs@knots=knots}. 

\noindent{\it Projecting discretized functional data.}
The function {\tt project()} works also when the input is a discretization of some continuous argument functional data.
This is probably the most important projection for functional data analysis. 
In this case, the input is a matrix having in the first column a set of arguments and in the remaining ones the corresponding values of a sample of functional data. 
The input data are considered to be piecewise constant functions with the value over two subsequent arguments equal to the value in the input corresponding
to the left-hand side argument.
In this way, the discretized data can be viewed as functions, and their inner products with any spline are well defined. 
In the package, this specific inner product can be obtained by utilizing the indefinite integral implemented in the function {\tt integra()}. 
Consequently, the projection $\mathbf P S$ of the functional data in $S$ satisfies (\ref{projection}), if $S$ represents the data as a  piece-wise constant function. 
In Figure~\ref{fig:Projectiion}, we illustrate the output from {\tt project()} for a sample of ten functional data that are projected to the space of splines generated by the spline in Figure~\ref{fig:splinets}~{\it (Left)}.

\section{Classification problem - an FDA workflow using {\tt Splinet}}
As we have seen, the package {\tt Splinet} provides a comprehensive toolbox to analyze functional data. 
However, for a specific goal, one has to form a workflow that assures all steps have been performed to form the final conclusions following the analysis. 
While the workflow may vary depending on the type of statistical analysis, some steps are essential regardless of the analysis type.
Here, we introduce a workflow for the classification problems.
The focus here is on functional data that have a one-dimensional domain. 
In the future, we plan to extend the analytical tools to functional data with domains in higher dimensions such as images, movies, etc. 
For this reason, we have chosen a classic data set of images; however, our approach to their two-dimensional nature involves converting them into functions of one argument.
Before introducing the actual workflow, we explain the fundamentals of our methodology, highlighting the innovations that are specific to our use of spline spaces. 
\subsection{Methodology}
In our approach to the classification of functional data, we search first for a proper functional space of splines that is capable of capturing important functional features of the data and, at the same, is thrifty as far as the dimension of the space is concerned. 
These two apparently opposite requirements need to be chosen in a balanced manner to facilitate both efficiency and precision. 
In the approach it is achieved by a smart choice of knots and the method is described in detail in \cite{basna2023empirically}.
Here, we only mention that the method places the knots in locations where the reduction in the total square error of data approximation by $0$-order splines is largest. 
The idea is basically the same as in building the regression tree but the criteria for obtaining the optimal selection are different. This is implemented in the {\tt ddk} package written in \textbf{\textsf{R}}, currently accessible on its GitHub page: \url{https://github.com/ranibasna/ddk}.

For classification, spline spaces are selected separately for each class.
Then one projects the training data into the respective spaces these functions from now on are referred to as functional data points. 
On these functional data points, the functional principal component analysis (FPCA) is performed on each class in the training set. 
That is, we calculate the means for each class. The so-centered data are decomposed into the eigenvalues and eigenfunction, i.e. we obtain spectral decomposition for each class separately. 
This involves estimating the eigenvalues ${\lambda_{i1}> \lambda_{i2} >\dots > \lambda_{in_i}}$ for the $i$th class and the corresponding set of eigenfunctions $\{e_{i1}, e_{i2},\dots, e_{in_i}\}$, where $n_i$ is the number of eigenfunctions per class $i$ and $ i=1,2, \dots, K$, where $K$ is the number of classes. To determine the optimal number of eigenvalues and eigenfunctions, we employ a validation method.
Theoretical validation of the spectral decomposition is through Karhune-Lo\`eve's representation of the functional data which states that a function in each class is sampled independently from the following model 
\begin{align}
\label{eq:spectdec}
X_i(t)=\mu_i(t) +\sum_{k=1}^\infty \sqrt{\lambda_{ki}} Z_{ki} e_{ki}(t),\, i=1,\dots, K,  
\end{align}
where $Z_{ki}$'s, mean-zero variance-one variables that are uncorrelated within a class and independent between classes. 

Subsequently, each original discretized data point, say $x_l$, from a test set is projected onto all $K$ spline spaces based on the original knot selection per class. 
Thus one obtains $K$ functional data points $f_{li}$, $i=1,\dots, K$. 
The core principle of classification is to determine how closely the $f_{li}$ test data point matches the projection to the eigenspaces for a given class. The mathematical techniques used in this process for image data include feature extraction, dimensionality reduction, and classification algorithms, which are used to obtain meaningful information from functional data points and classify them efficiently. 
Let us describe this classification procedure in some further detail.

The first step involves computing the mean for each of the ten classes using our training data sets. 
For a given class such a mean function of all elements in the training set belonging to the $i$th class is denoted by $\hat \mu_i$ and is viewed as an estimated value of $\mu_i$.
It is obtained by projection of the averaged data per class into a space of splines that has been determined by the knot selection process specifically for the mean.

For each original data point $x_l$ in the test set, one evaluates its representations $f_{li}$, $i=1,\dots, K$ in the knot selection driven spline spaces for each class.   
Our objective is to assess the proximity to each of the $K$ classes by projecting $f_{li}-\hat \mu_i$ onto the eigenspaces (the space spanned by eigenfunctions) corresponding to that class. 
Thus, we obtain $K$ distinctive projections. We denote them by $\hat{f}_{l1},\hat{f}_{l2}, \dots, \hat{f}_{lK}$. Explicitly,
\begin{equation}
\label{eq:eigennu}
 \hat{f}_{li} =\hat \mu_i + \sum_{j=1}^{n_i} \langle f_{li}-\hat\mu_i,\hat e_{ji} \rangle \hat e_{ji},\, i=1,\dots, K   
\end{equation}
where $\langle\cdot \rangle$ stands for the inner product in the functional spaces, i.e. integral over the product of two functions, and $\hat e_{ij}$ are estimates of eigenfunctions $e_{ij}$ obtained in the training phase and discussed above. 
Formally, we obtain the following spectral decompositions
$$
f_{li}=\hat f_{li} +\hat\varepsilon_{li},
$$
where $\hat\varepsilon_{li}$ is the residual of the projection.
If the functional point $x_l$ belongs to the $i$th class and $\hat \mu_i$ and $\hat e_{ij}$ are approximately equal to the true values, we would have based on \eqref{eq:spectdec}:
\begin{align*}
\hat\varepsilon_{li}
\approx 
\sum_{j={n_i}+1}^\infty \sqrt{\lambda_{ki}} Z_{ki}  e_{ji}(t),
\end{align*}
which should be rather small in the squared norm $\|\cdot\|$ of the functional spaces. 
Indeed, we have
$$
\|\hat\varepsilon_{li}\|^2\approx
\left \|
\sum_{j={n_i}+1}^\infty \sqrt{\lambda_{ki}} Z_{ki} e_{ji}
\right \|^2
=
\sum_{j=n_i+1}^\infty \lambda_{ki} Z_{ki}^2 
$$
 and thus 
 $$
 \mathbb E (\|\hat\varepsilon_{li}\|^2)
 \approx
 \sum_{j={n_i}+1}^\infty \lambda_{ki}
 ,
 $$ 
 which can be made small as long as the selection of $n_i$ targets values
\begin{align*}
 \sum_{j=1}^\infty \lambda_{ji} &\approx   \sum_{j=1}^{n_i} \lambda_{ji},
\end{align*}

On the other hand, if $x_l$ does not belong to the $i$th class, the residual $\hat\varepsilon_{li}$ should be large, given that the classes are sufficiently distinguished by projections to the eigenspaces of their largest eigenvectors.   
This justifies the following classification rule 
\begin{equation}
    \label{eq:class}
    I(x_l)=\mathop{\rm arg\, min}_{i=1,\dots,K}\|\hat \varepsilon_{li}\|=\mathop{\rm arg\, min}_{i=1,\dots,K}\|x_l -\hat f_{li}\|,
\end{equation}
where $I(x_l)$ stands for the chosen class and $x_l$ is treated as a piecewise constant function. 
One can enhance the outcome of classification by providing the squared normalized distances
\begin{equation}\label{eq:weights}
\left(w_1^l,\dots, w_K^l\right)=\frac{\left(\|x_l -\hat f_{l1}\|^2,\dots, \|x_l-\hat f_{lK}\|^2\right)}{\sum_{i=1}^K \|x_l-\hat f_{li}\|^2}.
\end{equation}

\begin{remark}
    This rule may be improved to account for the classes that differ not by eigenvectors but by the corresponding eigenvalues. 
    Namely, one could consider the path of residuals over an increasing number of eigenvectors and consider the sizes of the residuals along this path. 
    We do not investigate this enhancement of the proposed classification. 
\end{remark}

Having outlined the classification procedure, the pivotal role of selecting the appropriate eigenvalues/eigenfunctions for each class becomes evident. These hyperparameters have a great impact on our classification results.
To enhance our classification's precision, we ascertain the ideal number of eigenvalues or eigenfunctions. Using the validation set technique, our primary objective is to identify the optimal count of eigenvalues/eigenfunctions. 

\subsection{Workflow}
We describe, step-by-step, the workflow for the above-described methodology and, along with this, we perform these steps on the well-known Fashion MNIST dataset. The data set consists of a training set of 60,000 examples and a test set of 10,000 examples of Zalando's article images. The test dataset is divided into two distinct subsets: the validation dataset and the test dataset. Each data entry is represented by a grayscale image of $28 \times 28$ pixels. Items belong to one of 10 types of clothing, such as shoes, dresses, etc. The name of each class and its corresponding label are given in Table~\ref{class}. Thus the original problem can be viewed as a classification problem on the $784$-dimensional Euclidean space.  

\begin{figure}[t!]
  \centering
\raisebox{-0mm}{\includegraphics[width=0.32\textwidth]{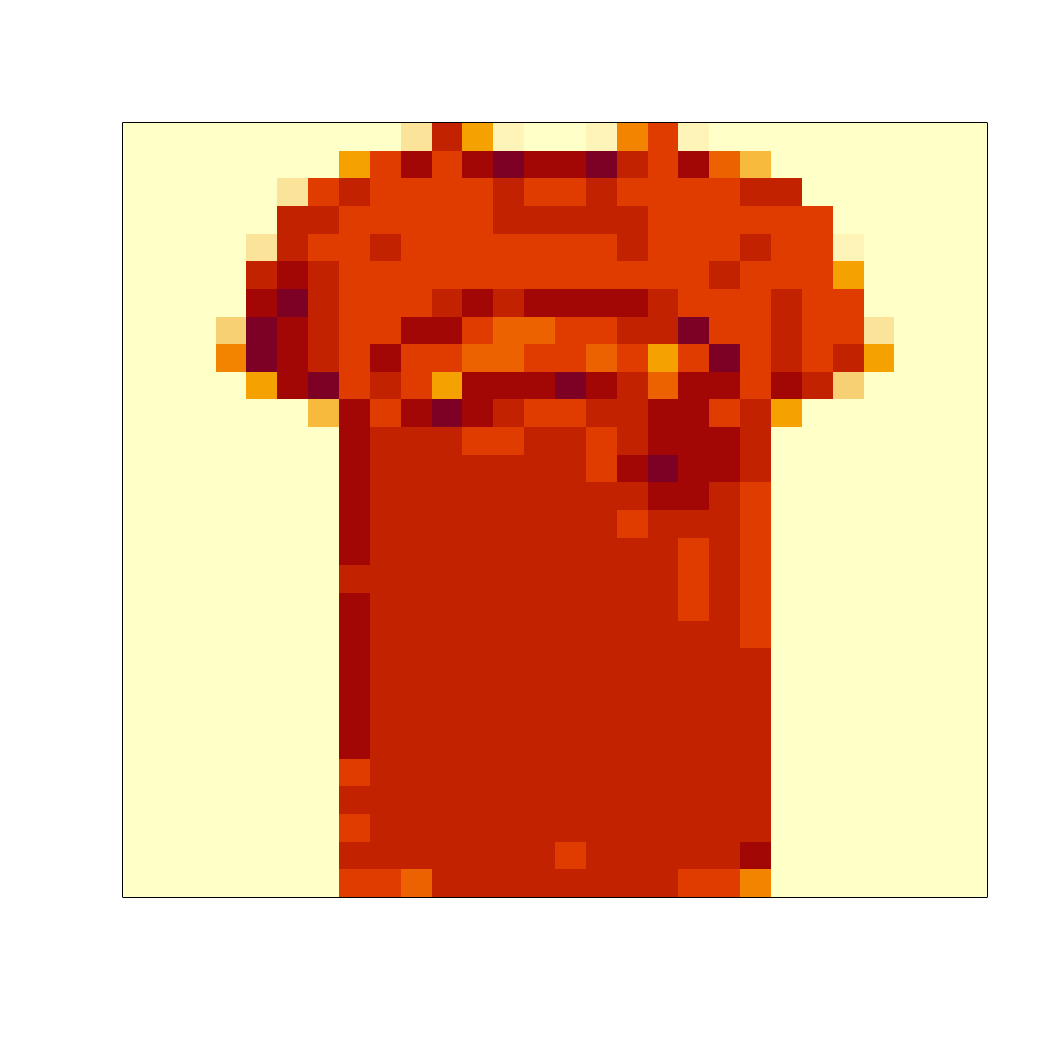}} \hspace{-2mm}
\includegraphics[width=0.33\textwidth]{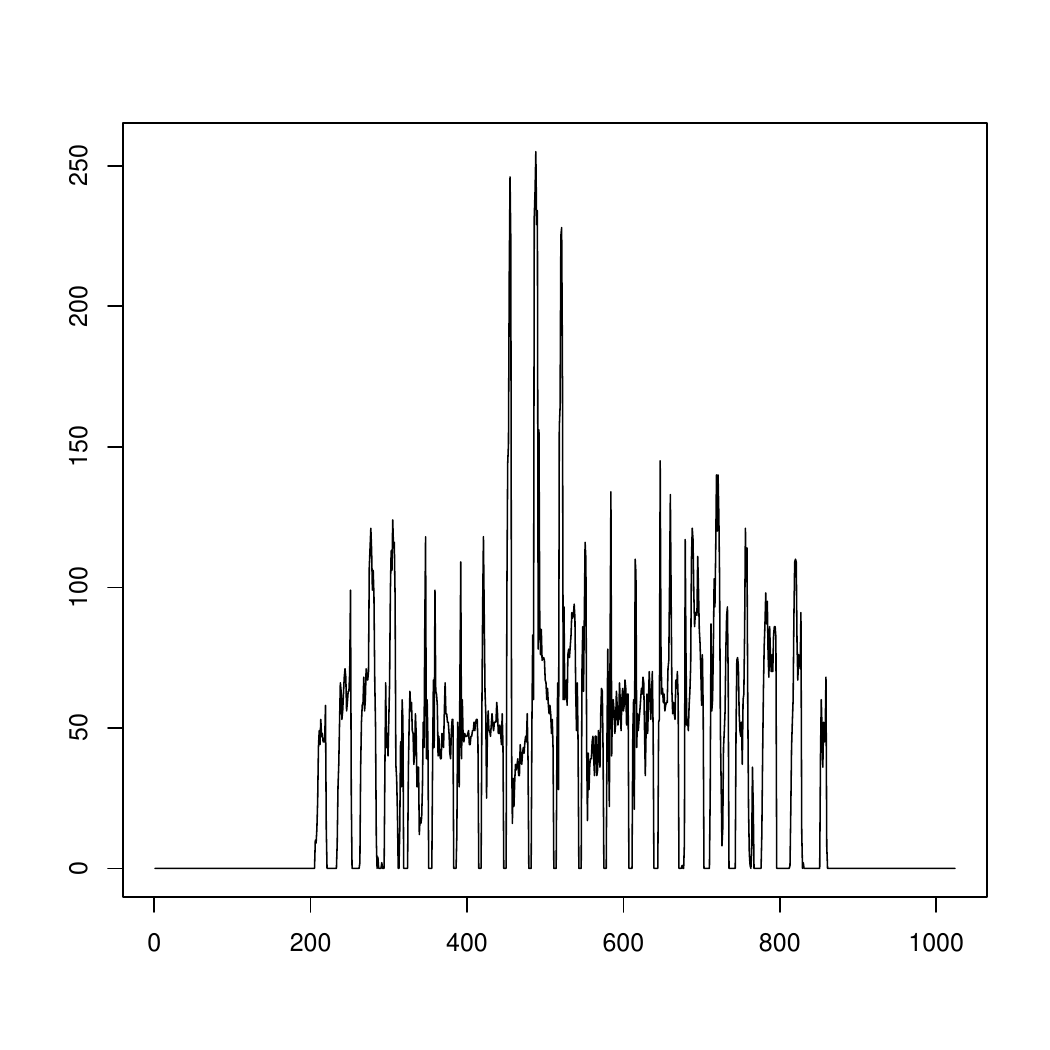}
\includegraphics[width=0.33\textwidth]{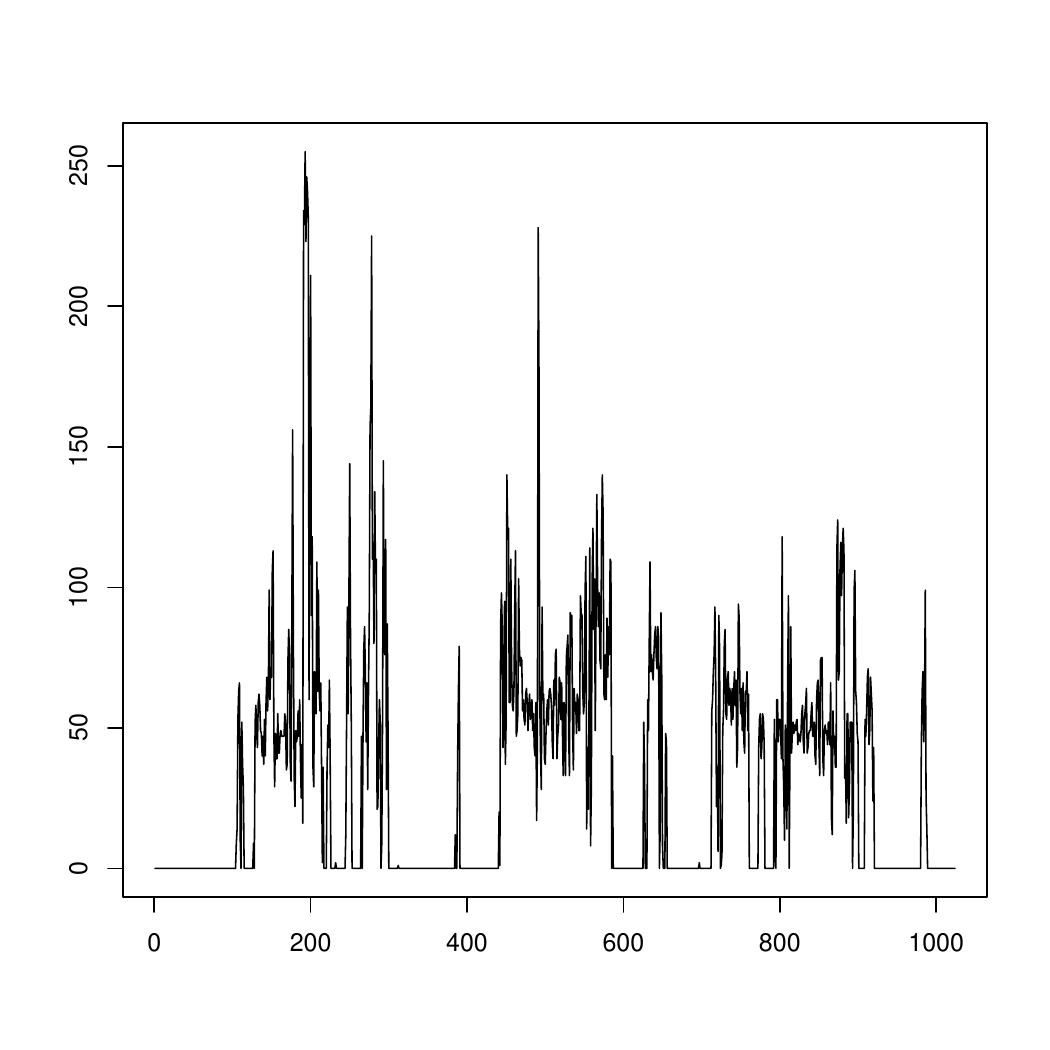}
\\
\vspace{-.3cm}
\raisebox{-0mm}{\includegraphics[width=0.32\textwidth]{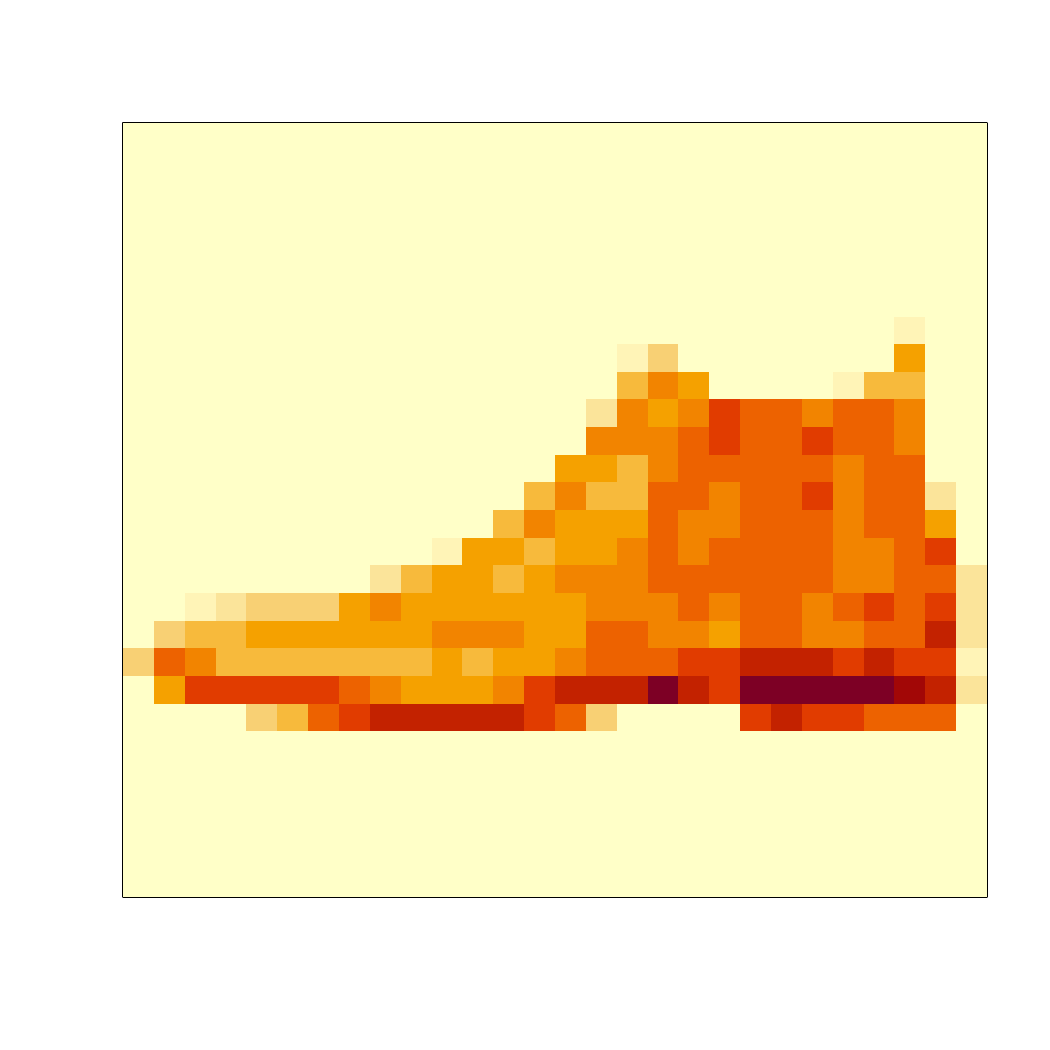} }\hspace{-2mm}
\includegraphics[width=0.33\textwidth]{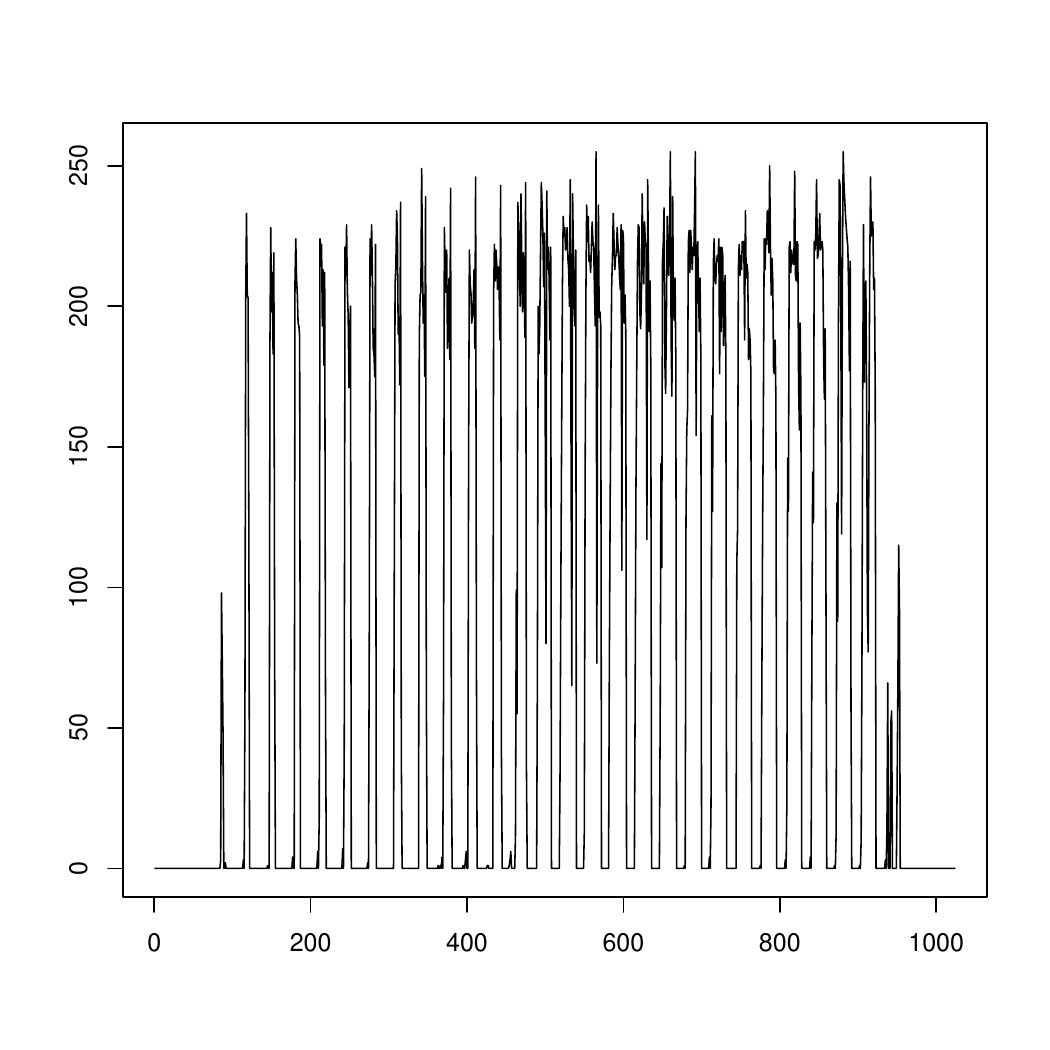}
\includegraphics[width=0.33\textwidth]{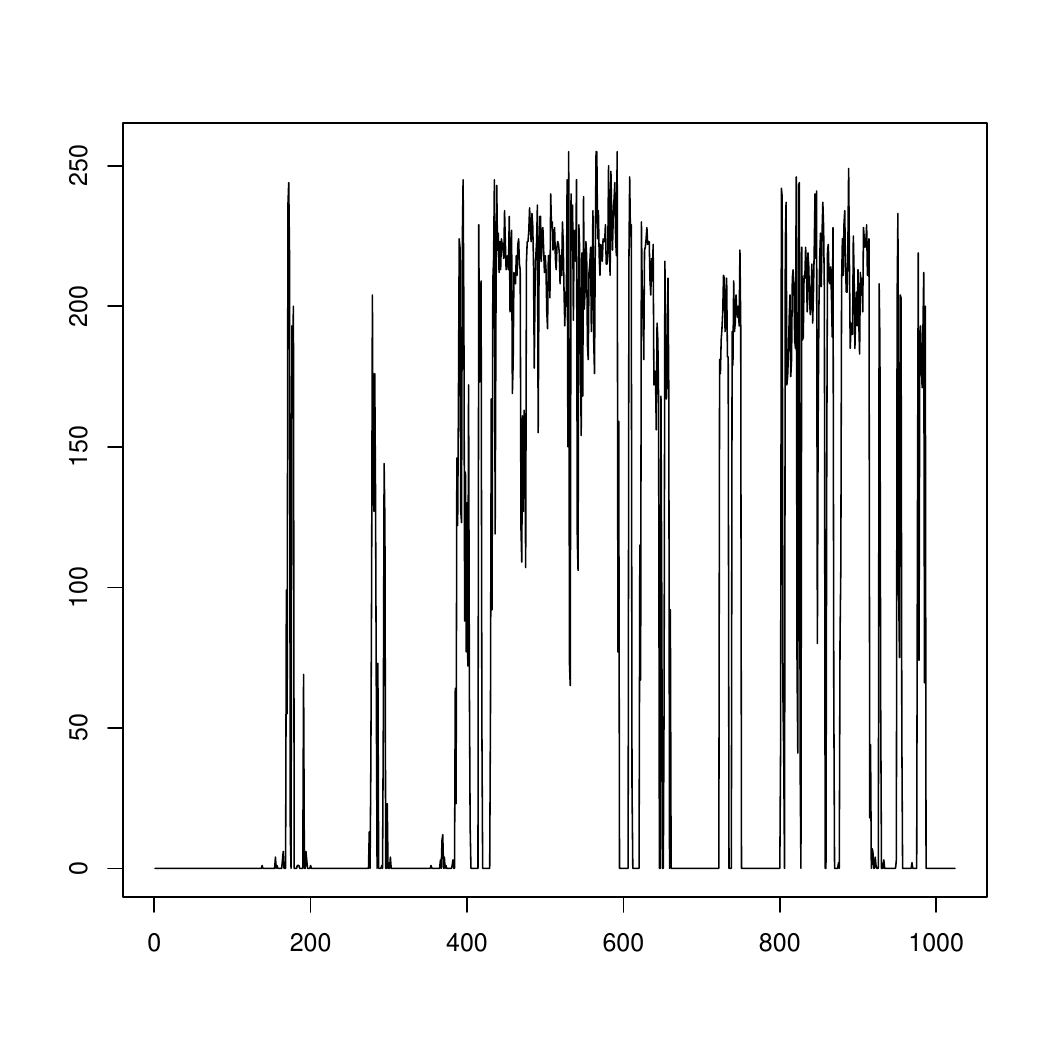}
  \caption{One- and two-dimensional representations of two examples of MNIST data, a T-shirt and a boot. {\it Left:} original images, 
  {\it Middle:} column-major order representation of the T-shirt and the boot. 
  {\it Right:} Hilbert curves representations of the two examples.}
  \label{HC}
  \end{figure}

\begin{table}[h]
\caption{Class names and labels of MNIST dataset.}
\centering
\small
\begin{tabular}{cccc}
\toprule Labels & Description & Labels & Description \\
\midrule 0 & T-shirt/top & 5 & Sandal\\
\midrule 1 & Trouser & 6 & Shirt \\
\midrule 2 & Pullover& 7 & Sneaker \\
\midrule 3 & Dress &8 & Bag \\
\midrule 4 & Coat &9 & Ankle boot\\
\bottomrule
\end{tabular}
\label{class}
\end{table}

\begin{enumerate}[leftmargin=0.4cm]
    \item{\bf Data preparation:} In this segment of the workflow, the package {\tt Splinet} is not used but the way we represent and preprocess our data plays a critical role in subsequent analyses and outcomes. 
    The remaining steps of the workflow assume that the data represent discretized functions, i.e. are matrices of columns representing arguments and values of the functional data. 
    Two variants of this format are allowed: one with the common arguments for all data points, where the vector of arguments stands as the last column; the second one allows different arguments for different data points, in which the input should be a list of two-column matrices, with the vector of the arguments as the first column and the vector of the corresponding values as the second one. 
    We note that in the second case, it is allowed to have a varying number of rows in the elements of the list. 

    Once the data are properly formatted, it should be divided into three parts. The first and largest part (typically at least $50\%$ of the data) constitutes the training data, and the remaining are split into two, the validation and testing. 
    Alternatively, one can perform training and cross-validation on a data set of size, for example, $75\%$, and test the approach on the remaining testing portion of the data.

    \begin{Example*}[Fashion MNIST dataset]
        For the Fashion MNIST dataset, we deal with two-dimensional images, and transforming the data to one dimension is a critical process in which information will be lost. One of the tasks, we settled with the analysis of this data set is to compare different approaches to this transformation process. For this reason, we elucidate our approaches to data representation and transformation.
 
Images are typically represented as matrices of pixels, where the value of each matrix element indicates the color and intensity of the respective part of the image.
The matrices need to be transformed into vector form. A prevailing approach for such a transformation involves stacking the image's columns (or rows) consecutively to create a vectorized representation. However, this technique often neglects the local spatial correlations existing between image pixels, ensuring only the preservation of vertical (or horizontal) correlations.

To more effectively retain these local correlations, we turn to the Hilbert curve transformation. A Hilbert curve is a continuous fractal space-filling curve that traverses every point in a square grid sized to any 2-power magnitude.  For a detailed discussion of Hilbert curves, we refer to \cite{bader2012space}. Hilbert curves are particularly interesting for their ability to group pixels locally. See the additional materials accompanying the paper for the \textbf{\textsf{R}}-code of the Hilbert space-filling algorithm.

Given that our image dimensions do not align with a 2-power magnitude, we have implemented zero padding, adjusting the image to a $2^5\times 2^5$ pixel matrices. Figure~\ref{HC} presents two illustrative examples from the MNIST dataset: a shirt and a boot. The middle illustrations provide vectorized representations of these items, derived from consecutively stacking image columns. The left-hand-side depictions visualize vectors formed through the Hilbert space-filling curve. It is easy to notice that the locality-preserving properties of Hilbert's curve make it a better choice in comparison to other options.

Another approach that can preserve more two-dimensional features of the images, when transformed to one dimension, is to consider gradient images, i.e. images in which the value of the discretized gradient is attached to each pixel. The gradient $G$ at the pixel $p=(i,j)$ of the image $I$ is defined as $G(p)=\sqrt{\Delta^2_{x,p}+\Delta^2_{y,p}}$, where $\Delta_{x,p}=\frac{I(i+1,j)-I(i-1,j)}2$ and $\Delta_{y,p}=\frac{I(i,j+1)-I(i,j-1)}2$. 
Thus, one has effectively two-dimensional vector valued data points $(I,G)$ that could be transformed into the one-dimensional domain through a Hilbert curve, and the one-dimensional FPCA can be applied to the transformed bivariate data. In our illustrative example, we do not follow this path.
    \end{Example*} 

\begin{figure}[t!]
\hspace{-8mm}
\raisebox{2mm}{\includegraphics[width=1.1\textwidth]{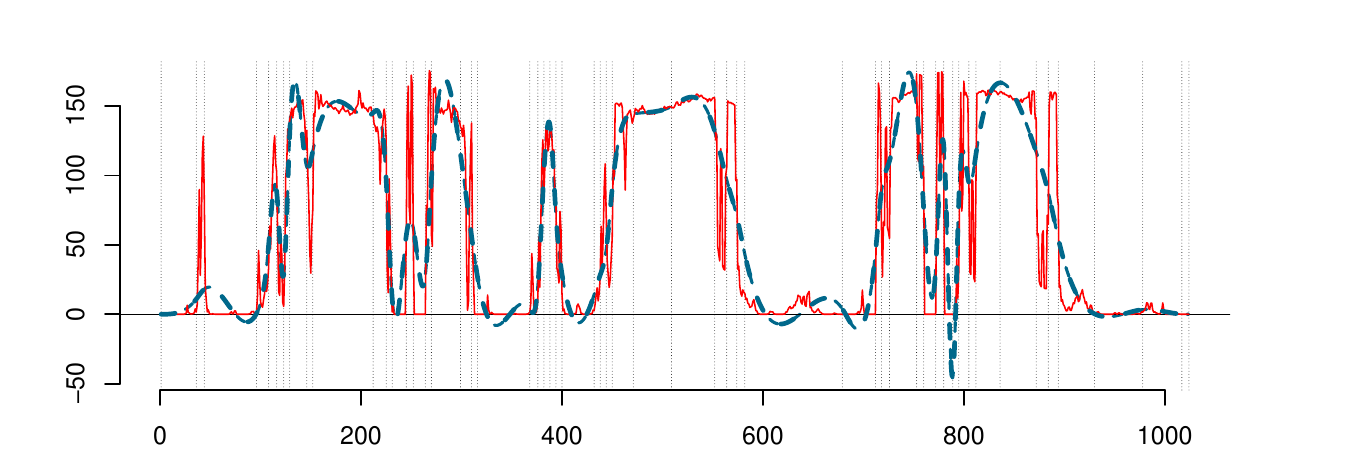}}  \hspace{-2mm}\vspace{-12mm}\\
\mbox{}\hspace{-8mm}
\raisebox{-0mm}{\includegraphics[width=1.1\textwidth]{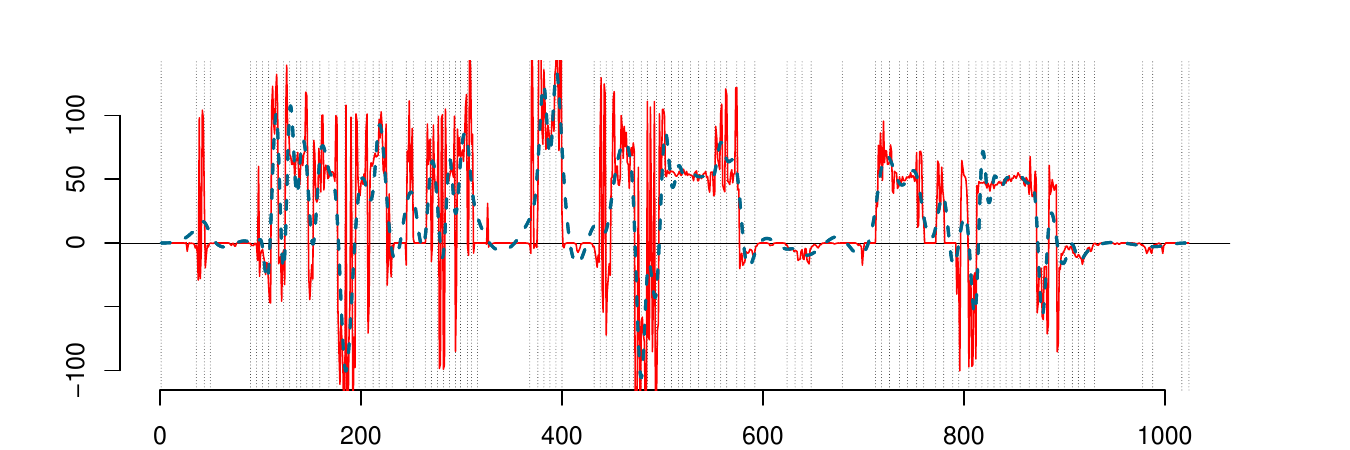}}\vspace{-12mm} \hspace{-2mm}\\
\raisebox{-0mm}{\includegraphics[width=0.5\textwidth]{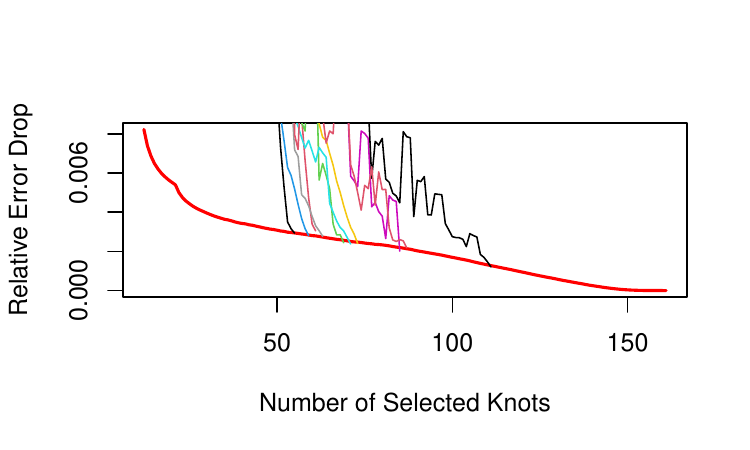}} \hspace{-2mm}
\raisebox{-0mm}{\includegraphics[width=0.5\textwidth]{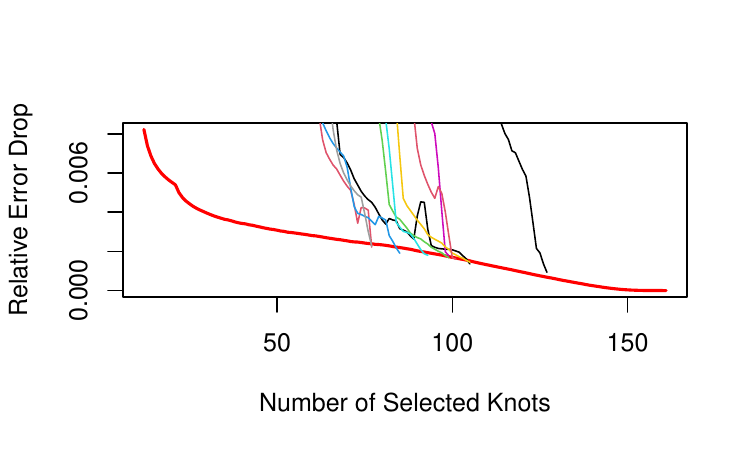}}\vspace{-3mm} \hspace{-2mm}
  \caption{
  The knot selection for the mean functions $\mu_k$, $k=1,\dots,K$ and the centered data.
  {\it Top:} Discrete mean `T-Shirt' (solid line) vs. the functional fit $\hat \mu$ (dashed line) obtained by the optimal knots marked by vertical dotted lines,
 {\it Middle:} Fitting the centered `T-shirt' by a spline with selected knots,
 {\it Bottom-Left:} Visualization of the stopping algorithm for the knots for the mean functions,
 {\it Right-Bottom:} Visualization of the stopping algorithm for the knot selection of the centered data. 
}
  \label{fig:KnotSelMean}
\end{figure}

    \item {\bf Data-driven knot selection:}
Selecting an appropriate basis for data representation as functional entities is one of the fundamentals of the proposed dimension reduction.
It facilitates the initial choice of a functional space that is appropriate for the data at hand.  
The Data-Driven Knots (DDK) selection algorithm, as detailed in \cite{basna2022data}, strategically places knots to capture the data's curvature. 
This is done to minimize the mean square error when projecting onto the space of piecewise constant functions for all samples in the training dataset within each class.
These selected knots are subsequently used to build a suitable basis for projecting the data. The algorithm has embedded a stopping criteria mechanism aimed at automating the convergence time of the knot selection.
Due to the novelty of this approach, we elaborate details of the proposed implementation. 

Since we assume \eqref{eq:spectdec}, it is clear that the mean value function $\mu$ and the eigenfunctions $e_i$, $i=1,\dots, \infty$ do not need to have anything in common.
For this reason, they should be treated separately. 
In particular, there are separate processes to select the knots. 
We start with the selection for the mean function. 
The first step is to calculate the mean of the data per class in the training data and then perform a knot selection for this estimated discretized mean.
Since we want to adopt an efficient stopping rule for the number of knots, we used the relative error reduction as a measure of reduction when adding a knot. 
Namely, if $AMSE_k(r)$ stands for the average mean square error of the approximation with $r$ knots for the mean in the $k$th class, or for fitting white noise if $k=0$, then we define the relative error drop by 
$$
\epsilon_k(r)=\frac{AMSE_k(r+1)-AMSE_k(r)}{AMSE_k(r)}, \, k=1,\dots,K,
$$
where $k=0$ corresponds to the relative error for the fitting noise and $k\ge 1$ for the knot selection of the mean in the class $k$.
The knots are stopped to be added if the relative error matches the drop of the adding knots to fit the noise, i.e. we adopt the following stopping rule
$$
N_k=\max\{r\ge 1, \epsilon_k(r)\ge \epsilon_{0}(r) \}.
$$
The $\epsilon_0(k)$ is the universal reference curve computed by the Monte Carlo method and independent of the specific case. 

Exactly, the same approach is chosen to select the knots for the centered data, except in this case, the total $AMSE_k(r)$ is evaluated across all samples from the given class in the training data. 
The starting knots for this part of the selection process are the knots obtained for the case of the mean. 
This assures that both mean and eigenfunction will reside in the same space of splines.  
The process signifies a mechanism through which the final optimal number $N_k$, $k=1,\dots,K$, of knots and their locations class is determined. 
The DDK algorithm is available as a package on the GitHub page: \url{https://github.com/ranibasna/ddk} and its function {\tt add\_knots(0)} is utilized.   

\begin{Example*}[Fashion MNIST dataset, cont.]
    In our process, the DDK algorithm is initially used in the training dataset to select knots for each of the ten classes. Consequently, the algorithm pinpoints recursively knots, placing them in regions exhibiting the most pronounced curvature within each class. 
    In Figure~\ref{fig:KnotSelMean}~{\it (Right-Top)}, we visualize the stopping rule for the ten classes yields values of the number of knots 
    $$
(N^0_1,N^0_2,N^0_3,N^0_4,N^0_5,N^0_6,N^0_7,N^0_8,N^0_9,N^0_{10})
=
(56,  62,  70,  60,  72, 86,  74,  64, 112, 88).
    $$ 
    The resulting functional version of the mean for 'T-shirts' is seen in Figure~\ref{fig:KnotSelMean}~{\it (Left-Top)} in a dotted line. 
    The resulting knot selection for the `T-shirt'-class
    can be seen in the locations of the dashed vertical lines. There is a clear correlation between the density of knots and the variability of a curve.

    Visualization of the stopping time for the number and location of knots for centered data is shown in Figure~\ref{fig:KnotSelMean}~{\it (Right-Bottom)}. 
    The algorithm has chosen the following number of knots for the respective classes
    $$
    (N_1,N_2,N_3,N_4,N_5,N_6,N_7,N_8,N_9,N_{10})=(106 , 78, 100 ,86, 94, 100, 106,  78, 128, 102).
    $$ 
    We observe a clear, although moderate increase in dimensionality in each of the classes. 
    The result suggests that the `Trouser' and `Sneaker' classes are the least complex, as they demand a space of lower dimensions (74-dimensions) while the `Bag' class requires higher dimensions (124-dimensions).
    Since the choices of knots are randomized, the dimension reduction should be considered as a very rough estimate of the complexity.
  In our example, we present the knot selection for the Hilbert curve approach only.

 Quantification of the importance of the knot selection is provided by the averaged weighted relative $L_2$-distance of the classes from their respective spaces, where the relative distance is taken as the $L-2$ distance divided by the sum of the distances of an element to other classes (this distances are used in the classification procedure described later). Here are the distances
 \begin{equation}
 \label{eq:reldist}
 (0.079, 0.103,  0.086,  0.089, 0.081, 0.085, 0.076, 0.085,  0.066, 0.073)
\end{equation}
 If the knot selection does not matter these distances should be around $0.1$ and, as we can observe, it happens only for the `Trouser' class.
 Notably, the `easiest' to represent is the `Bag' class. 
 \end{Example*}

\item {\bf Projection into spline spaces built on selected knots:} In this step, we consider $K$ distinct third-order data-driven splinet bases, each built on the knots specifically chosen for the corresponding class of the $K$ available. We then project the discrete training data points onto these splinets, constructed based on the class-specific knots.
We considered separated spaces for the mean discrete data and the centered discrete data. 
We note that since the knots for the mean data are a subset of the knots for the centered data, the projections of the means are in the space of the projections of the centered data. However, they are not the same as their projections of the centered data space. 

These projections establish an isomorphism between the Euclidean vectors made of coefficients of the projection and the space of splines into which the original data have been projected. 
More specifically, if $N_i$ is the number of knots chosen of the $i$th class and $k$ is the order of the splines, then the splines spanned on these knots constitute a $N_i-k-1$-dimensional Hilbert space with $N_i-k-1$ elements of the corresponding splinet so that 
\begin{equation}
\label{eq:isom}
x_l\stackrel{\tiny proj}{\longmapsto} f_{li} \stackrel{\tiny isom}{\longleftrightarrow} \mathbf a_{li}\in \mathbb R^{N_i-k-1},
\end{equation}
where $x_l$ is either the original centered discrete data point, $f_{li}$ is the corresponding functional data point when projected on the knots selected for the $i$th class, $\mathbf a_{li}$ is the vector of coefficients corresponding to the splinet expansion of $f_{li}$. 
This isomorphism allows us to perform the next part of the analysis simply by doing standard multivariate analysis on $\mathbf a_{li}$'s.
The projection process is facilitated using the {\tt project()} functions in the \textbf{\textsf{R}} {\tt Splinet} package.
Other functions of the package assist exploration and visualization of various features of $f_{li}$'s.
\begin{Example*}[Fashion MNIST dataset, cont.]
 The original images that were transformed through the Hilbert curve transform to the 1D discrete data are next projected to spline spaces. Considering the number of knots and since we consider the third order splines the dimension of the spline spaces are to be computed and put here according to the formula $(N_i-4)$. We note the initial dimension reduction from $784$ to the reported values of the given knot numbers in the classes. Additional further dimension reduction will be achieved through FPCA later on in the workflow.  We see the projection of the `T-Shirt' mean and an example of a  centered `T-Shirt' in Figure~\ref{fig:KnotSelMean}~{\it (Top-Middle)}. The locations of knots for these two cases are shown by vertical dashed lines.     
\end{Example*}

\begin{figure}[t!]
  \includegraphics[width=0.55\textwidth]{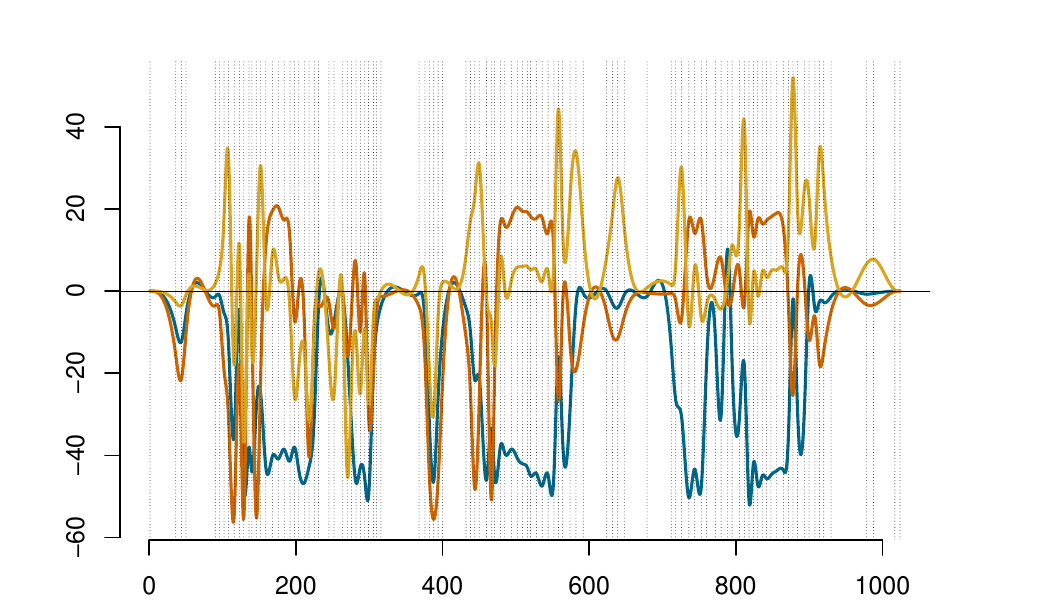}\hspace{-10mm}
    \includegraphics[width=0.55\textwidth]{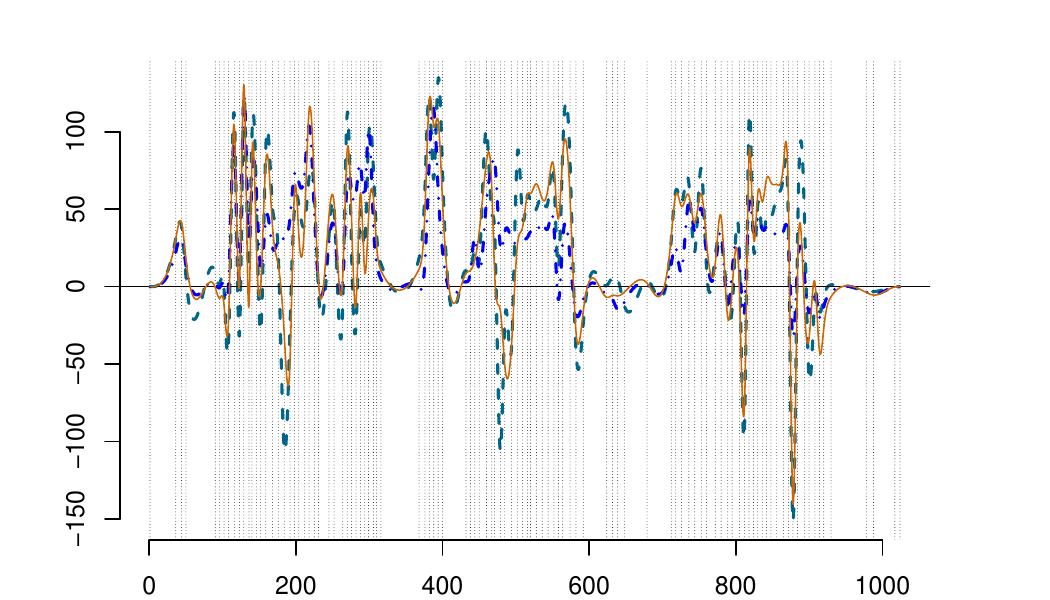}\hspace{-5mm}\\
  \includegraphics[width=0.55\textwidth]{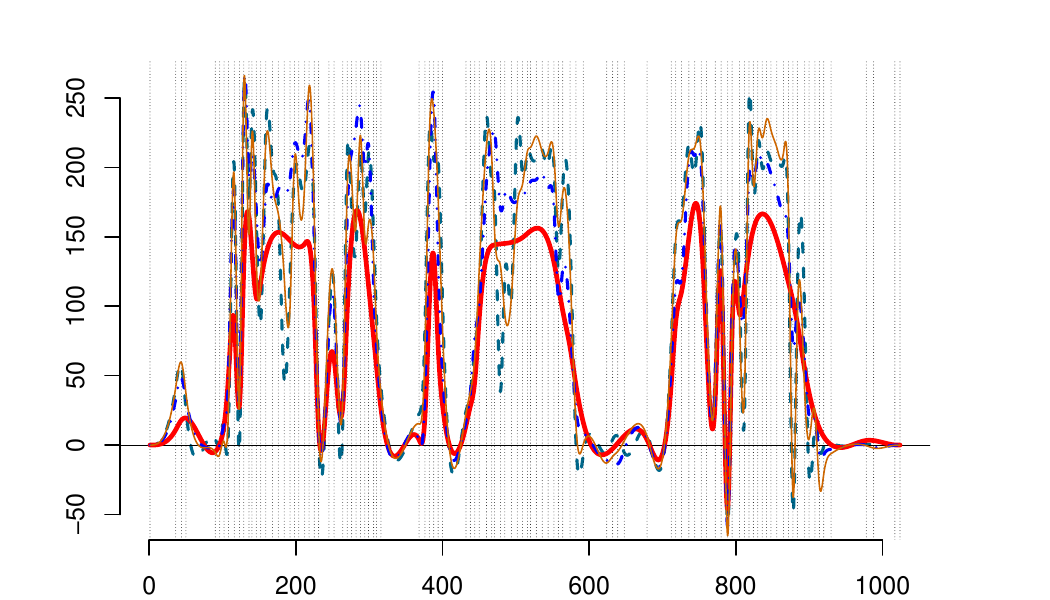}\hspace{-10mm}
  \includegraphics[width=0.55\textwidth]{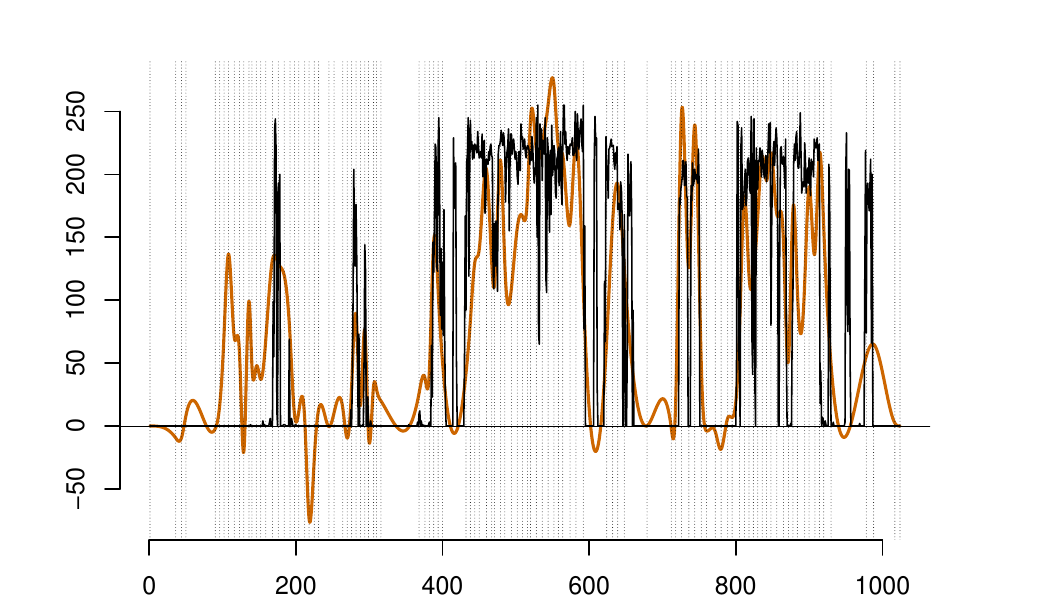}\hspace{-5mm}
  \caption{The spectral decomposition of the training data. 
  {\it Top-Left:} The first three eigenfunctions for the `T-shirt' class scaled by the square roots of the respective eigenfunctions, 
  {\it Top-Right:} Two approximations of the centered functional `T-shirt' data point {\it (NavyBlue-Dashed Line)}: 1) by the first three eigenfunctions {\it (Blue-DottedDashed Line)} ; 2) by the first twenty eigenfunctions {\it (Orange-ThinSolid Line)},
  {\it Bottom-Left:}
  The same approximation but centered around the `T-Shirt' class mean $\hat \mu_1$ {\it (Red-ThickSolid Line)},
  {\it Bottom-Right:}
  Projection of a 'Boot' data point to the 'T-Shirt' spectrum with 20 eigenfunctions after centering around the 'T-Shirt' class mean, the discrete`Boot' data point {\it (Black-Rough Line)} vs. the projection {\it (Orange-Solid Line).}
}
  \label{KS_EigenFunctions}
  \end{figure}
 
\item {\bf FPCA on training data:}
 To facilitate the classification procedure in the next step but also to gain a deeper understanding of the complexity of the data, the FPCA is performed within each class of the functional data points of the training set. In this task, one utilizes the isometry given in \eqref{eq:isom}, and first the sample covariance matrix $\boldsymbol \Sigma_i$ is evaluated
 $$
 \boldsymbol \Sigma_i = \overline{\mathbf a_{.i}\mathbf a_{.i}^\top} -\overline{\mathbf a_{.i}}\,\,\overline{\mathbf a_{.i}}^\top,
 $$
 where averaging $\overline{\cdot}$ is made within the class $i$ of the training data. 
 For this in \textbf{\textsf{R}}, one simply applies {\tt cov()} on the matrix of column vectors $\mathbf a_{li}$'s. 
 Then the spectral decomposition of $\boldsymbol \Sigma_i$ into eigenvalues and eigenvectors is performed using {\tt eigen()} of the \textbf{\textsf{R}} {\tt base}-package. 
 The obtained eigenvectors correspond to eigenfunctions through the isometry.

 \begin{Example*}[Fashion MNIST dataset, cont.]
    This rather standard step when performed on cloth image classes is illustrated in Figure~\ref{KS_EigenFunctions}. The first three eigenfunctions of `T-shirts' scaled by the respective eigenvalues are shown in the {\it (top-left)} graph and projections to the spaces based on 3 and 20 eigenfunctions are presented in the remaining graphs including the projection of  `Boots' to the `T-Shirt' space, the {\it (bottom-right)} graph. 
 \end{Example*}

\item {\bf Determining the significant eigenfunctions:}
In this step, the classification procedure \eqref{eq:class} as a function of the number of considered eigenfunctions is implemented.
Then the number $n_i$ of the eigenfunctions for the $i$th class is based on its accuracy on the validation data set.
In Figure~\ref{KS_EigenFunctions}~{\it (Bottom)}, we see the illustration of the classification principle. 
There are shown approximation of two data points, a `T-Shirt' {\it (Left)}, and a `Boot' {\it (Right)}, based on the projection to 20 eigenvalues in the 'T-Shirt spline space. It can be seen clearly that the approximation of 'T-shirt' is better than the approximation of `Boot'. 
The functional $L_2$-norm of the functional spaces can be utilized to measure the approximation. 

The approach to data classification relies on projecting the data into a subspace defined by a set of numbers $n_i$, $i=1,\dots, K$, of eigenfunctions of the distinct classes.
We can see from \eqref{eq:eigennu} and the classification rule \eqref{eq:class} that if $n_i$'s are taken too big, then it may result in overfitting of the data and a smaller dimension reduction, on the other hand, if the values are too small the precision of distinguishing the features in the data of different classes may be not sufficient. 
In this framework, the numbers of eigenfunctions for individual classes are of paramount significance and are treated as hyperparameters.
The $n_i$'s, $i=1,\dots,K$ are chosen through the following cross-validation procedure. 

From the training data, from each class, we exclude at random $10\%$-data points, denoted by $\mathcal C_i$, $i=1,\dots,K$. 
Consider the remaining $90\%$ for the $i$th class and build spectral decomposition of the data as explained in the previous step.
Set the initial vector of numbers of the eigenvalues $\mathbf n^0=(n^0_1,\dots,n_{10}^0)$ in each class. For example, we can set it to zero to start with the classification based only on the mean $\hat \mu_i$, i.e. the closest mean decides for the class to which a data point belongs to. 
Let the classification distances (evaluated through \eqref{eq:weights}) for a given discrete data point $x$ be denoted by 
$$
\mathbf w(x,{\mathbf n^0})=\mathbf w(x;n^0_1,\dots,n_{10}^0).
$$
Evaluate classification success rate through
$$
\mathbf s({\mathbf n^0})=\left(\frac{\sum_{x\in \mathcal C_1} w_1(x,\mathbf n^0)}{|\mathcal C_1|},\dots,\frac{\sum_{x\in \mathcal C_K} w_K(x,\mathbf n^0)}{|\mathcal C_K|} \right).
$$
where $|\mathcal C_i|$ is the number of elements in $\mathcal C_i$, or through the classification accuracy rate
$$
\mathbf a({\mathbf n^0})=\left(\frac{\#\{x\in \mathcal C_1; I(x,\mathbf n^0)=1\}}{|\mathcal C_1|} ,\dots,\frac{\#\{x\in \mathcal C_K; I(x,\mathbf n^0)=K\}}{|\mathcal C_K|} \right).
$$
Small values in $\mathbf s$ or large values in $\mathbf a$ indicate good classification. 
Thus, taking averages of these two vectors can be used to assess the overall quality of a classification rule. 
Since the accuracy $\mathbf a$ is more interpretable, we focus on it and denote the average of its entries by $\bar {\mathbf a}$. 

In general, the function $\mathbf n\mapsto \bar {\mathbf a}(\mathbf n)$ is a non-convex function of $K$ arguments, and our goal is to find its optimum when evaluated over the validation data set.  
We adopt a simple iterative marginal gradient method, where we increase by one that coordinates in $\mathbf n$ that produces the large increase in $\bar {\mathbf a}$. 
Allow for a specific number of negative increases $L$ and conclude with the location $\mathbf n_{opt}$ of the maximum over-searched path.  
The number $L$ is a hyperparameter and is data-specific. 
The algorithm is repeated by a number of random initial knot distributions and the final result is the maximum over these runs.
 \begin{figure}[t!]
  \centering
\includegraphics[width=0.5\textwidth]{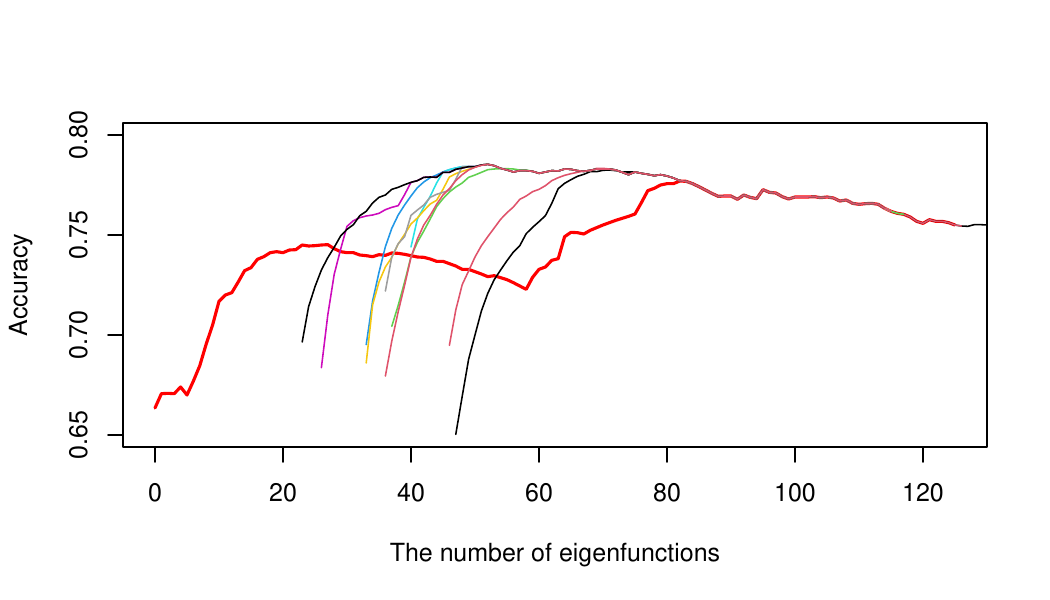}\hspace{-2mm}
\includegraphics[width=0.5\textwidth]{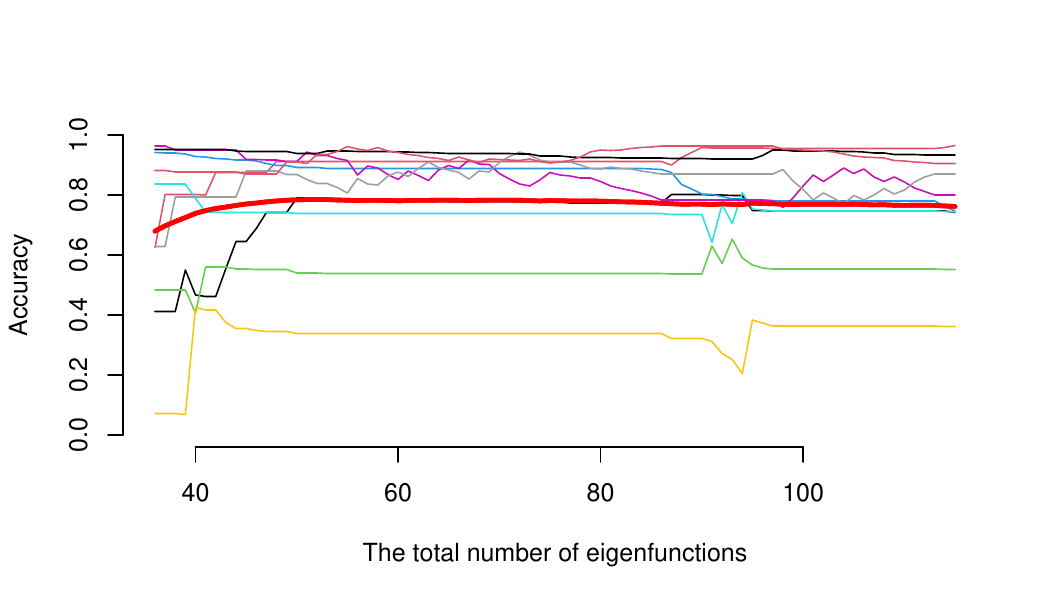}

  \caption{Optimization of accuracy in the validation phase (Step 5). {\it Left:} The trajectory of the average accuracy along the initial optimization path (thick-line) together with subsequently chosen random samples of the initial $\mathbf n_0$, {\it Right:} The class-wise accuracies against the average accuracy. }
  \label{VAL}
  \end{figure}
\begin{Example*}[Fashion MNIST dataset, cont.] The procedure was run on the validation set of the functional using the classification procedure \eqref{eq:class} (also used in the next step of the workflow). 
The baseline for the performance is made of the rates of correct classification per class when it is only based on the distance of a data point from the spline projection of the mean, i.e. $\|x-\hat \mu_i\|$, so no eigenfunctions are considered. 
The obtained rates of correct classification rates per class are 
$$
(0.67, 0.88, 0.33, 0.73, 0.66, 0.75, 0.23, 0.80, 0.76, 0.86)
$$
with the overall average performance of $66\%$ of correct classification.
From this, one sees that `Pullover' and `Shirt' seem to be difficult to classify.

In Figure~\ref{VAL}, the trajectories of the classwise accuracies and the average accuracies based on the above optimization procedure are presented. They lead to
the following numbers of significant eigenvectors $(  8,  4,  5,  6,  4, 8,  2, 4,  5, 6)$ for the ten cloth classes and, with this choice, the average accuracy in the validation process is $78.5\%$. We conclude that the original dimension of $784$ vectors has been reduced more than tenfold (based on the total number $52$ of all eigenfunctions). 
However, we also see that `Pullovers' and `Shirt' remain hard to classify, with the latter being properly classified less than $40\%$ of the time. 
The final classification based on the obtained sizes of the functional spaces is performed on the testing data set in the next step. 
\end{Example*}

\item {\bf Testing the classification procedure:}
This step is essentially repeating on the testing data set, most of the workflow except the validation steps used for the knot selection and deciding for the number of eigenfunctions. 
Thus for each data point $x_l$ in the testing sets, one projects it to ten functional spaces obtaining splines $f_{li}$, $i=1,\dots, K$. Those in turn are used to evaluate the classification rule \eqref{eq:class} and decide on the class of the object. 
In fact, in the algorithm, the following convenient representation of the squared distance between a discrete data point $x_l$ and its approximation $\hat f_{li}$ given in \eqref{eq:eigennu} is used 
$$
\|x_l-\hat f_{li}\|^2=\|x_l\|^2-\|f_{li}\|^2 + \|f_{li}-\hat \mu_i\|^2
-\sum_{j=1}^{n_i}\left| \langle f_{li}-\hat \mu_i,\hat e_{ji}\rangle\right|^2.
$$
The results are compared with actual class memberships and summarized in the confusion matrix and other standard measurements of efficiency. 

\begin{Example*}[Fashion MNIST dataset, cont.] 
The leading example illustrating the workflow is using Hilbert curve approach to transform from two-dimensional images to one-dimensional discrete data points.  
The classification method \eqref{eq:class} with optimally chosen numbers of eigenvectors was performed through all testing data points and several characteristics have been evaluated.
The average accuracy (over all classes) is $77.0\%$, which is close to the one in the cross-validation step and the classwise accuracies are 
$$
(73.4\%, 89.1\% ,52.3\%, 87.7\%, 72.0\% , 91.7\%, 32.9\%, 82.2\%, 94.5\%, 92.9\%),
$$
which are also consistent with the results in the cross-validation experiment seen in Figure~\ref{VAL}~{\it (Right)}. 

Another important summary is to report the averaged normalized distances $\bar{\mathbf w}$, of the elements of the class to their projection to that class and they are
$$
 (0.049, 0.049, 0.053, 0.046, 0.047, 0.051, 0.060, 0.033, 0.042, 0.030).
$$
We observe expected negative correlations with the accuracies. The values are significantly lower from $0.1$ (the case that the distance does not differentiate between classes) and are a major improvement over the original relative distances to the spaces given in \eqref{eq:reldist}. 
\end{Example*}
\item {\bf Final evaluation and conclusions:}
This part depends on the goal for which the classification procedure has been used and thus is case-specific. 
One may need simply a classification method and its accuracy, then little is needed beyond the details shared in the previous step.
    It is also recommended to examine the statistics of misclassified data points and the features of these points as expressed by FPCA. 
    This may give insight into the reasons for failing to identify data points properly and, in consequence, give an idea of how the classification can be further improved. 
    If classification needs to be applied to some data without labels, then the classification should be run on it and summarized. 
    This workflow can also serve as a benchmark to compare various classification methods.
   Then a comparison and a discussion should be carried out through reporting confusion matrices, along with other pertinent evaluation metrics. Again checking which data points have been misclassified can be important to see if a hybrid classification method could further improve the success rate.
\begin{Example*}
In our illustrative example, the goal was to show the workflow and the corresponding methodology. 
Since we focused on a single method, it is natural to present here more detailed characteristics of the method and comment on the obtained outcomes.
It is clear that the method performs poorly in classifying `Shirts' and, somewhat better, `Pullovers'.  We observe that a `Shirt' is often classified as a `T-Shirt' and a `Coat'. Additionally, a `Pullover' is often classified as a `Coat'. 
Generally, it seems that the main responsibilities for the misclassifications are the classes: `T-shirt', `Pullover', `Coat', and `Shirt'. 

To improve on the method one could try to use a lower order of splines that could help due to the noisiness of the Hilbert curve data points. 
By just looking at the data, it appears that the first-order splines should suffice.
Improving the search for an optimal number of eigenfunctions could be another possible source of improvement. 
We have seen that the performance from the validation step carries over to the testing results.
Thus, enhancing the validation process should lead to more accurate classification results.
Nevertheless, it seems that the achieved performance will be hard to significantly improve unless the full 2D character of the data is considered. 
This is planned in a future 2D extension of the spline-based method for the FDA. 

\begin{center}
\begin{table}[h]
\caption{Confusion matrix}
\centering
\footnotesize
 \begin{tabular}{l r r r r r r r r r r}
 \toprule 
 \multicolumn{11}{c}{\bf TARGET}\\
 \bf PREDICT. &T-shirt & Trouser & Pullover & Dress & Coat & Sandal & Shirt & Sneaker &Bag &Boot\\
 \midrule
T-shirt& 73.4\% &0.8\% &4.5\% &4.6\% &0.8\% &0.0\% & 25.5\% &0.0\% &2.2\% & 0.1\%\\
Trouser& 0.1 & 89.1\% &0.3\% &1.7\% &0.1\% &0.0\% &0.0\% &0.0\% &0.0\% & 0.0\%\\
Pullover& 1.8\% &0.5\% & 52.3\% &1.5\% & 10.4\% &0.0\% &8.6\% &0.0\% &0.7\% & 0.0\%\\
Dress & 12.1\% &8.7\% &2.0\% & 87.7\% &8.4\% &0.0\% &8.9\% &0.0\% &1.2\% & 0.0\% \\
Coat & 0.3\% &0.6\% & 20.2\% &1.7\% & 72.0\% &0.0\% & 17.9\% &0.0\% &0.2\% & 0.0\%\\
Sandal & 0.7\% &0.0\% &0.0\% &0.1\% &0.0\% & 91.7\% &0.0\% &9.2\% &1.0\% & 3.3\%\\
Shirt & 5.0\% &0.1\% & 17.5\% &1.8\% &6.8\% &0.0\% & 32.9\% &0.0\% &0.0\% & 0.0\%\\
Sneaker & 0.0\% &0.0\% &0.0\% &0.0\% &0.0\% &4.7\% &0.0\% & 82.2\% &0.1\% & 3.7\% \\
Bag & 6.6\% &0.2\% &3.0\% &0.8\% &1.4\% &0.4\% &6.2\% &0.0\% & 94.5\% & 0.0\% \\
Boot & 0.0\% &0.0\% &0.2\% &0.1\% &0.1\% &3.2\% &0.0\% &8.6\% &0.1\% &92.9\%\\
\bottomrule
\end{tabular}
\label{tab:confusion}
\end{table}
\end{center}

\end{Example*}
\end{enumerate}
\section{Comparing different 1D-methods functional method for Fashion-Mnist}
In this section, we employ the proposed workflow to investigate how different methods of data preparation influence the efficiency of our classification method.
Our main goal is to assess the impact on the efficiency of two factors: the Hilbert curve-based transformation method and the DDK selection. 
We put forward three distinct scenarios to be examined using our established workflow:
\begin{itemize}
    \item [S1:] Hilbert curve transformation with $100$ data-driven knot selections.
    \item [S2:] By-row data with $100$ data-driven knot selections.
    \item [S3:] By-row data with $100$ equidistant knots.
\end{itemize}
For the first two scenarios, we utilized the DKK algorithm \cite{basna2022data}.  Even though DDK can determine the optimal number of knots, we've consciously opted for a consistent count of 100 knots across all three scenarios. This choice ensures both the validity and fairness of our comparisons, thus enhancing the reliability and interpretability of our results. 

\begin{figure}[t!]
  \centering
\hspace*{-.8cm}\includegraphics[width=0.34\textwidth]{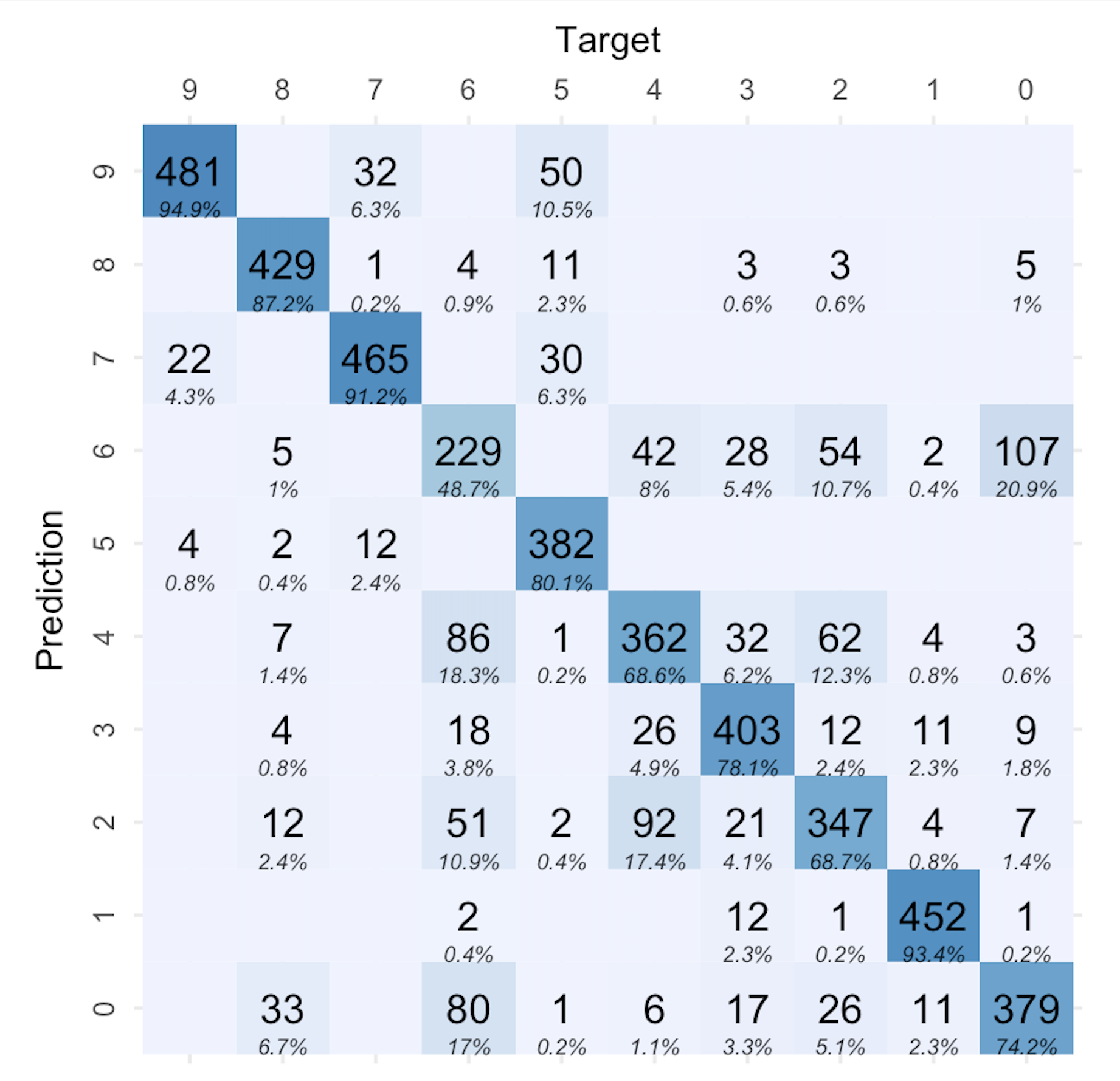}\includegraphics[width=0.36\textwidth]{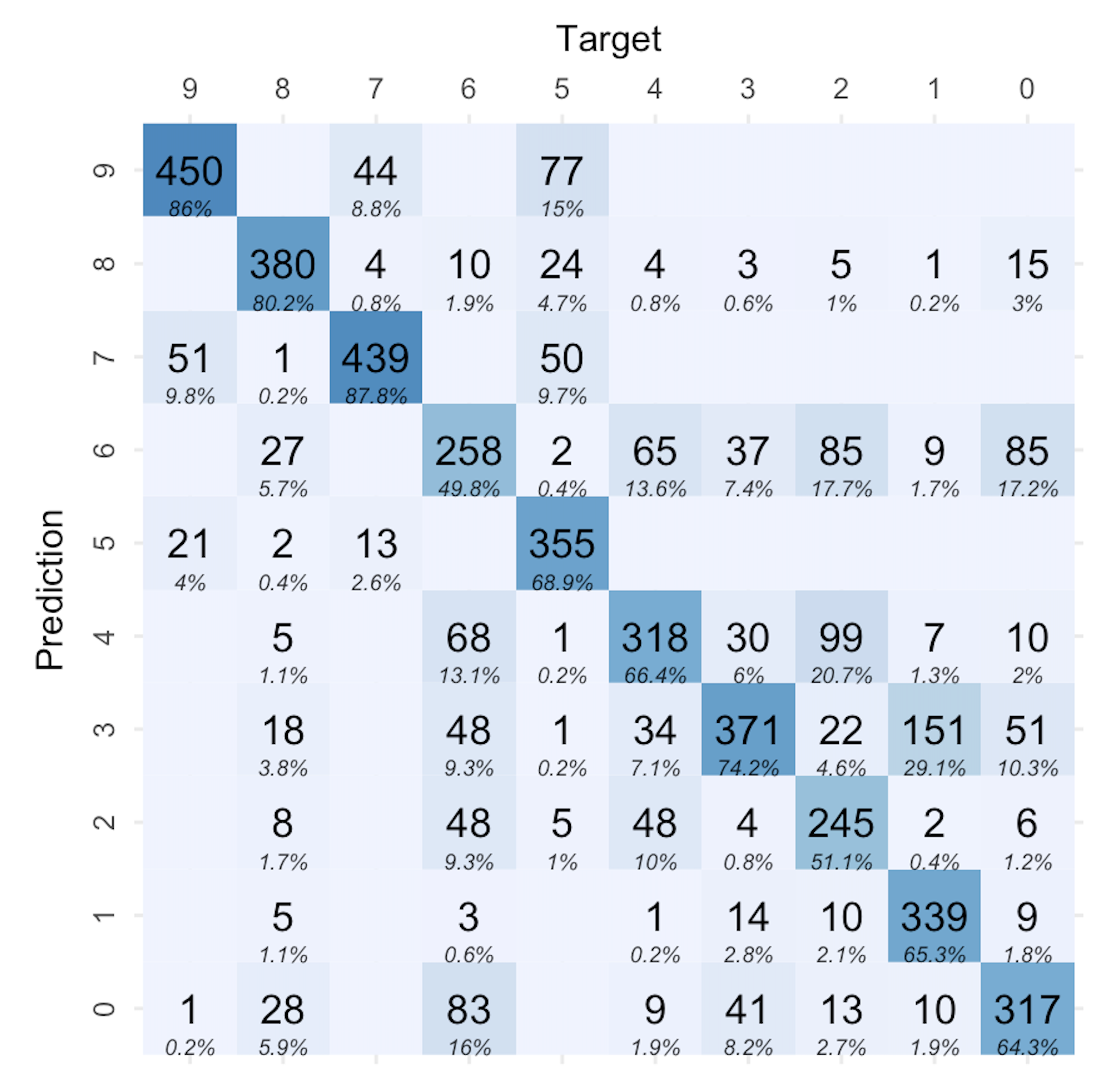}\includegraphics[width=0.36\textwidth]{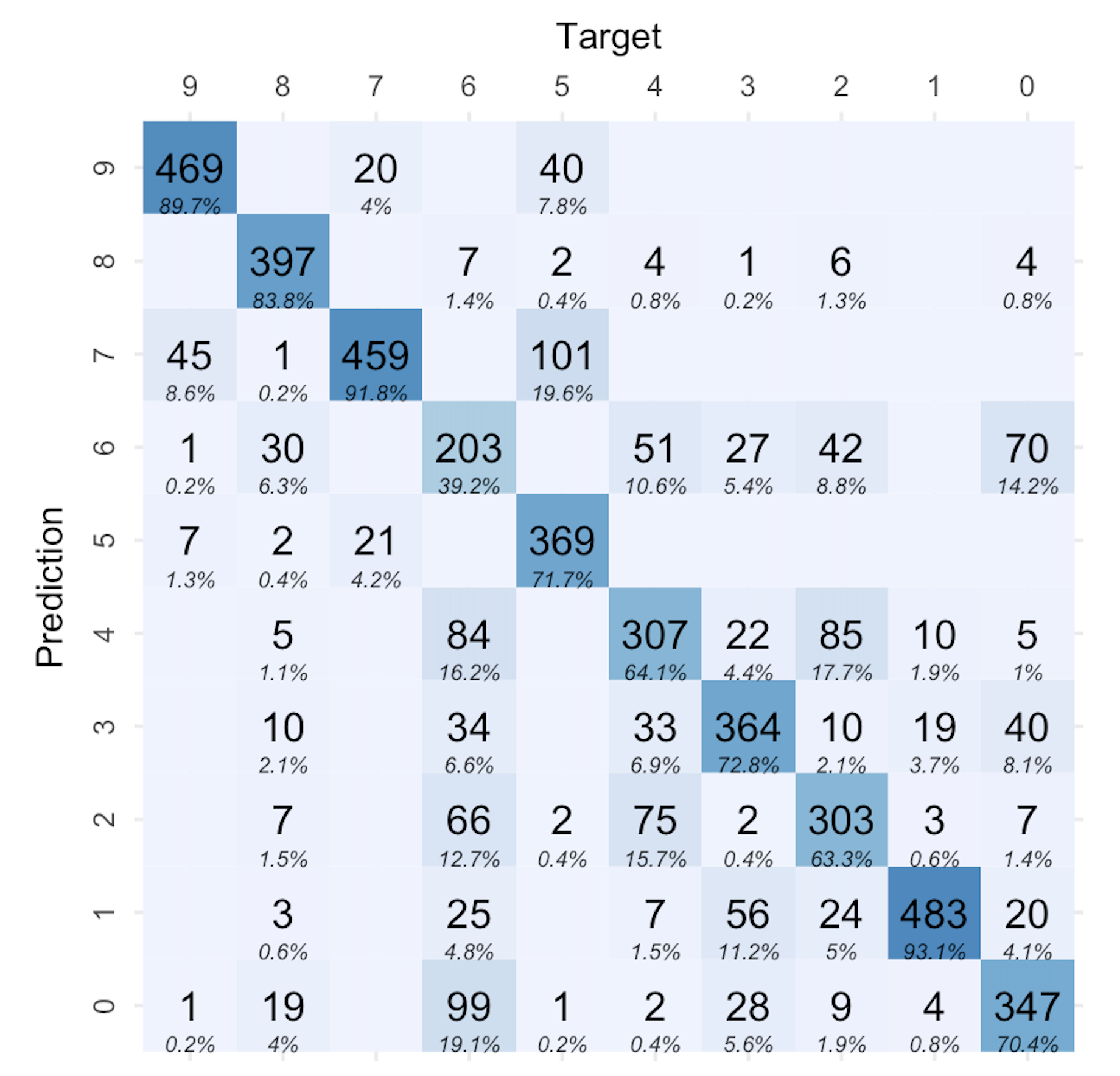}
  \caption{Confusion matrix for S1, S2, and S3 from right to left.}
  \label{CM}
  \end{figure}
  
The choice of these scenarios stems largely from our belief that our DDK selection method would be more compatible with the FPCA, leading to a more efficient dimension reduction, especially in the context of sparse data. As evidenced in Figure~\ref{HC}, the pixel distribution in the by-row data does not exhibit sparsity. This suggests that we might not see significant performance gains if we were to deploy the DDK knot selection methodology on it.
However, the introduction of the Hilbert curve transformation imparts a noticeable sparsity in the data's curvature. 
Thus it could be expected that this transformation could amplify the efficacy of the FPCA when integrated with the DDK approach, potentially leading to more accurate classifications.
As detailed above, we project the data into a subspace spanned by a number of eigenfunctions. For each scenario, we chose a number of eigenfunctions that achieved the highest accuracy on the validation data set. The validation phase of the analysis suggests that the optimal number of eigenvalues for S1, S2, and S3 are 15, 10, and 15, respectively. After the testing phase, we carry the classification problem described earlier across the three scenarios.
To evaluate the performance of our classification model, we define the following metrics: Accuracy, Precision, Recall, and F1 Score. The definitions of these metrics are as follows: 
\begin{equation*}
        \text{Accuracy} = \frac{TP + TN}{TP + TN + FP + FN}, \ \ \ \  \text{Precision} = \frac{TP}{TP + FP}, \ \ \ \  \text{Recall} = \frac{TP}{TP + FN}, 
        \end{equation*}
        where $TP, TN, FP, FN$ denote true positives, true negatives, false positives and false negatives, respectively. 
 The F1 Score is the harmonic mean of Precision and Recall, ideal for uneven class distributions:
        \begin{equation*}
        \text{F1 Score} = 2 \times \frac{\text{Precision} \times \text{Recall}}{\text{Precision} + \text{Recall}}.
\end{equation*}


Table \ref{3scenarios} presents the outcomes obtained from analyzing the classification problem within the three suggested scenarios.
\begin{center}
\begin{table}[h]
\caption{Accuracy, Precision, Recall and F1 Score of the classification problem in each scenario.}
\centering
\small
 \resizebox{0.7\columnwidth}{!}{
\tiny
 \begin{tabular}{c c c c } \toprule
  & Scenario 1 & Scenario 2 & Scenario 3\\ \hline 
 Accuracy&  $78.6\%$ & $69.44\%$ & $73.6\%$ \\ 
 Precision&  $79.22\%$ & $71.3\%$ & $74.1 \%$ \\
 Recall&  $78.5\%$ & $69.4\%$ & $73.9\%$ \\
 F1 Score&  $78.74\%$ & $69.8\%$ & $73.6\%$ \\
\bottomrule
\end{tabular}
}
\label{3scenarios}
\end{table}
\end{center}
The results from the first scenario (S1) clearly demonstrate its superior performance, affirming our proposition. To further understand the algorithm's misclassifications, we provide the confusion matrix for each scenario in Figure \ref{CM}. The confusion matrix illustrates the comparison between the true labels and the predicted labels of the test dataset. The confusion matrices show the uncertainty mostly between the classes 0,2,4,6 in the fashion image dataset, as discussed in the previous section . This makes sense because t-shirts, pullovers, coats, and shirts look similar and might be confusing.
 

 In the absence of the Hilbert curve transformation, it is observed that the classification performance of the data-driven knot selection approach alone is the least accurate among the three scenarios. This is consistent with expectations given that the data exhibits a dense and non-sparse format with highly repetitive patterns.
 The equidistant knot selection classification strategy shows a modest increment in the accuracy results. However, the application of data-driven knot selection on the transformed data significantly improves the accuracy of the results. It is important to note that our aim is not to outperform or compete with state-of-the-art machine learning and deep learning methodologies in image classification. 
 Although our method may not achieve the the optimal accuracy levels that can be seen in the convolution neural network approach, it does reach a commendable accuracy rate of approximately $80\%$. This is comparable to simpler neural networks, such as decision tree classifiers and MLP classifiers. For further benchmarking of machine learning algorithm classifiers on the Fashion MNIST dataset, see \cite{xiao2017fashion}. The implementation of the Hilbert curve transformation highlights the effectiveness of the data-driven method for representing data in a functional format, resulting a noticeable increase in accuracy by $5-6\%$. This enhancement is attributed to the efficient representation of locality features in the data through the chosen basis, ensuring the preservation of space curvature.

There are two notable observations to consider. First, the data-driven knot selection method demonstrates superior performance compared to the traditional equidistance method, especially when handling sparse data that exhibits distinct spatial dependencies. This difference in performance can be observed by comparing the class-wise accuracy between S1, S2, and S3. Second, leveraging the 2D characteristics of the images has a positive impact on the performance of Functional Principal Component Analysis (FPCA) when combined with the data-driven knot selection method. A transformation such as the Hilbert curve, can reveal the two-dimensional characteristics of the original image by preserving the locality relationships among the pixels. More precisely, in the flattened one-dimensional vector, consecutive elements (pixels) were likely neighbors in the original two-dimensional space. It also creates a sparser representation of the image, which makes it more amenable to our method. Therefore, when we apply our FPCA data-driven analysis techniques to this one-dimensional vector, these techniques approximately analyze the local neighborhoods of the original two-dimensional image. This enhancement suggests, as expected, that considering the 2D structure of the images contributes to improved results and should be further investigated.
\section{Summary}
We have presented a workflow for analyzing functional data using {\tt Splinet} and some auxiliary numerical tools, such as data-driven knot selection. 
This workflow is designed to enhance the efficiency of dimension reduction while retaining functional dependencies in the data's characterizing features, which may not necessarily be local. We applied this workflow to the benchmark Fashion MNIST dataset.
The findings from this example emphasize the pivotal importance of the data's two-dimensional nature in the functional principal component analysis approach. 
The Hilbert curve proved to be the most efficient and it is partially due to its capacity of capturing dependencies beyond one dimension. 
These observations suggest that adopting the method further toward two-dimensional analysis, for example, by considering gradient images or, even better, two-dimensional FPCA may lead to significant gains in the efficiencies.
This underscores the necessity to evolve the workflow towards two-dimensional extensions of the method. In the future, we aim to augment our data-driven FPCA methodology to account directly for 2D images with techniques that preserve the data and enhance the sparsity.

\bibliographystyle{unsrtnat}
\bibliography{RJreferences}

%% file: RJwrapper.bbl
\begin{thebibliography}{16}
\providecommand{\natexlab}[1]{#1}
\providecommand{\url}[1]{\texttt{#1}}
\expandafter\ifx\csname urlstyle\endcsname\relax
  \providecommand{\doi}[1]{doi: #1}\else
  \providecommand{\doi}{doi: \begingroup \urlstyle{rm}\Url}\fi

\bibitem[Perperoglou et~al.(2019)Perperoglou, Sauerbrei, Abrahamowicz, and Schmid]{Perperoglou:2019aa}
Aris Perperoglou, Willi Sauerbrei, Michal Abrahamowicz, and Matthias Schmid.
\newblock A review of spline function procedures in r.
\newblock \emph{BMC Medical Research Methodology}, 19\penalty0 (1):\penalty0 46, 2019.
\newblock \doi{10.1186/s12874-019-0666-3}.
\newblock URL \url{https://doi.org/10.1186/s12874-019-0666-3}.

\bibitem[de~Boor(1978)]{Boor1978APG}
Carl de~Boor.
\newblock A practical guide to splines.
\newblock In \emph{Applied Mathematical Sciences}, 1978.

\bibitem[Schumaker(2007)]{schumaker2007spline}
Larry Schumaker.
\newblock \emph{Spline functions: basic theory}.
\newblock Cambridge University Press, 2007.

\bibitem[Nguyen(2015)]{nguyen2015construction}
Tung Nguyen.
\newblock Construction of spline type orthogonal scaling functions and wavelets.
\newblock Honors Project Paper 19, Illinois Wesleyan University, 2015.

\bibitem[Goodman(2003)]{goodman2003class}
Tim~NT Goodman.
\newblock A class of orthogonal refinable functions and wavelets.
\newblock \emph{Constructive approximation}, 19\penalty0 (4):\penalty0 525--540, 2003.

\bibitem[Cho and Lai(2005)]{cho2005class}
Okkyung Cho and Ming~Jun Lai.
\newblock A class of compactly supported orthonormal b-spline wavelets.
\newblock \emph{Splines and Wavelets}, pages 123--151, 2005.

\bibitem[Liu et~al.(2022)Liu, Nassar, and Podg{\'o}rski]{LIU2022}
Xijia Liu, Hiba Nassar, and Krzysztof Podg{\'o}rski.
\newblock Dyadic diagonalization of positive definite band matrices and efficient b-spline orthogonalization.
\newblock \emph{Journal of Computational and Applied Mathematics}, 414:\penalty0 114444, 2022.
\newblock ISSN 0377-0427.
\newblock \doi{https://doi.org/10.1016/j.cam.2022.114444}.
\newblock URL \url{https://www.sciencedirect.com/science/article/pii/S0377042722002102}.

\bibitem[Mason et~al.(1993)Mason, Rodriguez, and Seatzu]{mason1993orthogonal}
JC~Mason, Giuseppe Rodriguez, and Sebastiano Seatzu.
\newblock Orthogonal splines based on b-splines -- with applications to least squares, smoothing and regularisation problems.
\newblock \emph{Numerical Algorithms}, 5\penalty0 (1):\penalty0 25--40, 1993.

\bibitem[Basna et~al.(2022)Basna, Nassar, and Podg{\'o}rski]{basna2022data}
Rani Basna, Hiba Nassar, and Krzysztof Podg{\'o}rski.
\newblock Data driven orthogonal basis selection for functional data analysis.
\newblock \emph{Journal of Multivariate Analysis}, 189:\penalty0 104868, 2022.

\bibitem[Nassar and Podgórski(2023)]{nassar2023splinets}
Hiba Nassar and Krzysztof Podgórski.
\newblock Splinets 1.5.0 -- periodic splinets, 2023.

\bibitem[Basna et~al.(2023)Basna, Nassar, and Podgórski]{basna2023empirically}
Rani Basna, Hiba Nassar, and Krzysztof Podgórski.
\newblock Empirically driven spline bases for functional principal component analysis in 2d, 2023.

\bibitem[Qin(2000)]{Qin}
K.~Qin.
\newblock General matrix representations for $b$-splines.
\newblock \emph{Vis. Comput.}, 16:\penalty0 177--186, 2000.

\bibitem[Zhou et~al.(2008)Zhou, Huang, and Carroll]{Zhou}
L.~Zhou, J.Z. Huang, and R.J. Carroll.
\newblock Joint modeling of paired sparse functional data using principal components.
\newblock \emph{Biometrika}, 95:\penalty0 601--619, 2008.

\bibitem[Redd(2012)]{Redd}
A.~Redd.
\newblock A comment on the orthogonalization of $b$-spline basis functions and their derivatives.
\newblock \emph{Stat. Comput}, 22:\penalty0 251--257, 2012.

\bibitem[Bader(2012)]{bader2012space}
Michael Bader.
\newblock \emph{Space-filling curves: an introduction with applications in scientific computing}, volume~9.
\newblock Springer Science \& Business Media, 2012.

\bibitem[Xiao et~al.(2017)Xiao, Rasul, and Vollgraf]{xiao2017fashion}
Han Xiao, Kashif Rasul, and Roland Vollgraf.
\newblock Fashion-mnist: a novel image dataset for benchmarking machine learning algorithms.
\newblock \emph{arXiv preprint arXiv:1708.07747}, 2017.

\end{thebibliography}
